\documentclass[useAMS,usenatbib,babel,times]{mn2e}
\usepackage{multirow,balance}
\usepackage[english,english]{babel}
\usepackage{amsmath}
\usepackage{amssymb,amsfonts,textcomp}
\usepackage{array}
\usepackage{supertabular}
\usepackage{hhline}
\usepackage{hyperref}
\usepackage[usenames]{color}
\hypersetup{dvips, colorlinks=true, linkcolor=black, citecolor=black, filecolor=black, urlcolor=black}
\usepackage[dvips]{graphicx}

\def\gtrsim{\lower.5ex\hbox{$\; \buildrel > \over \sim \;$}}
\usepackage{graphicx}

\newcommand{\msun}{\mbox{$M_\odot$}}

\newcommand{\hagn}{\mbox{{\sc \small Horizon-AGN\, }}}

\newcommand{\der}{{\rm d}}

\definecolor{grey}{rgb}{0.75,0.75,0.75}
\definecolor{Orange}{rgb}{1.0,0.5,0.15}
\definecolor{brown}{rgb}{0.7,0.25,0.0}
\definecolor{Pink}{rgb}{1.0,0.5,0.5}
\definecolor{darkerred}{rgb}{0.8,0,0}
\definecolor{darkerblue}{rgb}{0,0,0.8}
\definecolor{Blue}{rgb}{0,0.08,0.65}
\definecolor{Red}{rgb}{0.65,0.08,0.05}
\definecolor{Green}{rgb}{0.15,0.45,0.25}

\begin{document}

\author[Codis et al]{
\parbox[t]{\textwidth}{S.~Codis$^{1,2}$\thanks{E-mail: codis@iap.fr},
 R.~Gavazzi$^{1,2}$, Y.~Dubois$^{1,2}$, C.~Pichon$^{1,2}$, \\
  K.~Benabed$^{1,2}$,  V.~Desjacques$^{3}$, D. Pogosyan$^{4}$, J. Devriendt$^{5}$, A. Slyz$^{5}$ }
\vspace*{6pt}\\
\noindent$^{1}$CNRS, UMR7095, Institut d'Astrophysique de Paris, 98 bis Boulevard Arago, 75014, Paris, France\\
$^{2}$ Sorbonne Universit\'es, UPMC Univ. Paris 06, UMR7095, Institut d'Astrophysique de Paris, 98 bis Boulevard Arago, 75014, Paris, France\\
$^{3}$ D\'epartement de Physique Th\'eorique. Universit\'e de Gen\`eve, 24, quai Ernest Ansermet. 1211, Gen{\`e}ve, Switzerland\\
$^{4}$ Department of Physics, University of Alberta, 11322-89 Avenue, Edmonton, Alberta, T6G 2G7, Canada\\
$^{5}$ Sub-department of Astrophysics, University of Oxford, Keble Road, Oxford OX1 3RH\\
}


\title[Intrinsic alignment of galaxies ]{
Intrinsic alignment of simulated galaxies in the cosmic web:\\ implications for weak lensing surveys}

\maketitle

\begin{abstract}
{The intrinsic alignment of galaxy shapes { (by means of their angular momentum}) and their cross-correlation with the surrounding dark matter tidal field are investigated using the 160 000, $z=1.2$ synthetic galaxies extracted from the high-resolution cosmological hydrodynamical simulation {\sc \small Horizon-AGN}. One- and two-point statistics of the spin of the stellar component are measured as a function of mass and colour. For the low-mass galaxies, this spin is {\sl locally} aligned with the  tidal field `filamentary' direction while, for the high-mass galaxies, it is {\sl perpendicular}  to both filaments and walls. The bluest galaxies of our synthetic catalog are more strongly correlated with the surrounding tidal field than the reddest galaxies, and this correlation extends up to $\sim 10$ $h^{-1}\,\rm Mpc$ comoving distance. We also report a correlation of the 
projected ellipticities of blue, intermediate mass galaxies on a similar scale  at a level of $10^{-4}$ which could be a concern for cosmic shear measurements. We do not report any measurable 
intrinsic alignments of the reddest galaxies of our sample. This work is a first step toward the use of very realistic catalog of synthetic galaxies to evaluate the contamination of weak lensing measurement by the intrinsic galactic alignments.}
\end{abstract}


\section{Introduction}
For the last two decades, weak gravitational lensing has emerged as one of the most promising cosmological probes of the Dark Matter and Dark Energy contents of the Universe, culminating in the design of several large surveys like DES\footnote{http://www.darkenergysurvey.org}, Euclid
 \citep{Euclid} or LSST\footnote{
\texttt{http://www.lsst.org}}.

As the statistical power of weak lensing surveys is ramping up, more and more attention has to be paid for the control of systematic effects. Among the critical astrophysical sources of errors is the problem of the intrinsic alignments (IA) of galaxies. The fundamental assumption upon which galaxies are randomly aligned in the absence of a shear signal that is coherent on the scales of several arc minutes, is likely to break down for pairs of galaxies observed at close angular distances (through direct gravitational interactions or as a result of the same local tidal field they live in). 
Much effort has thus been made to control the level of IA of galaxies as a potential source of systematic errors in weak gravitational lensing measurements~\citep[e.g.][]{HRH00,C+M00,H+S04} although some techniques have been proposed to mitigate their nuisance by making extensive use of photometric redshifts \citep[eg][]{B+K07,J+S08,J+S10,J+B10,KBS10,Bla++12}.

Direct measurements of the alignment of the projected light distribution of galaxies in wide field imaging data seem to agree on a contamination at a level of a few percents in the shear correlation functions, although the amplitude of the effect depends on the depth of observations (stronger for shallower surveys), the amount of redshift information and the population of galaxies considered (in the sense that red galaxies seem to show a strong intrinsic projected shape alignment signal whereas observations only place upper limits in the amplitude of the signal for blue galaxies) \citep{Bro++02,L+P02,B+N02,Hey++04,Hir++04,Man++06,Hir++07,Man++11,Joa++11,Joa++13a}. Direct observations of the alignment between the spin and the tidal tensor eigenvectors have also been carried out: the first attempt by \citet{L+P02} studied the correlations between the disc orientation of the galaxies from the Tully catalogue and the shear reconstructed from the Point Source Catalogue Redshift survey and confidently rejected the hypothesis of randomness. More recently, \citet{leeetal13} detected some correlations between the spin and the intermediate eigenvector of the tidal tensor and found that galactic spins were also preferentially perpendicular to the major principal axis but this signal remains weak.

Cosmological numerical simulations are a natural way of further refining our models of IA. The imprint of the large-scale dynamics onto the shapes and spins of galaxies has been extensively studied using dark matter simulations \citep[among others]{aubertetal04,bailin&steinmetz05, calvoetal07, hahnetal07b, pazetal08, sousbie08,Lee++08,codisetal12,laigle2014,Forero-Romero2014}. One can also mention a numerical study of the alignment between halo spin and tidal tensor by \citet{porcianietal02} who predicted its orthogonality with the major principal axis but also found that galactic spins must have lost their initial alignment with the tidal tensor predicted by Tidal Torque Theory (TTT) and by \citet{codisetal12}. However, given the complex dependency on the physical properties of the galaxies seen in the observation, it is probably difficult to rely on dark matter-only numerical simulations as the sole resort to predict and control IA for weak lensing applications despite some success with the addition of halo model or semi-analytical models~\citep[e.g.][]{S+B10,Joa++13b}. 

\begin{figure*}
\includegraphics[width=2\columnwidth]{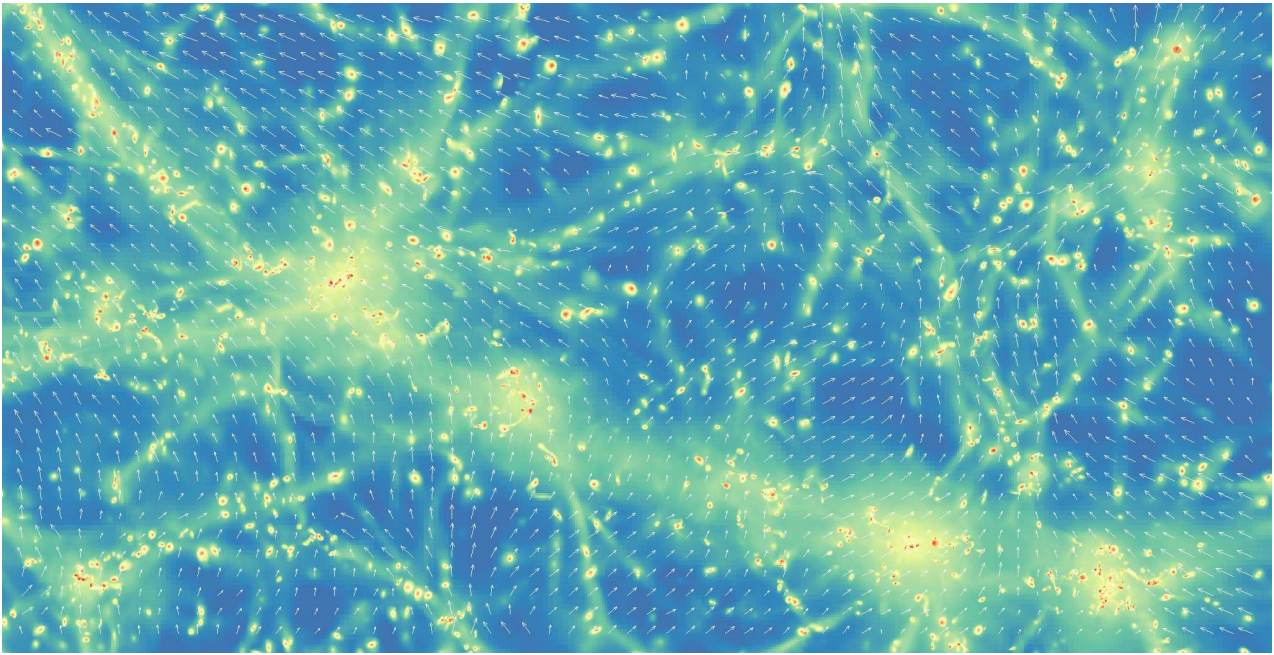}
\caption{{ The $\mathbf{e}_1$ eigenvector (white arrows) of the tidal field within a slice of 25 $h^{-1}\,\rm Mpc$ comoving in depth and $12.5$ $h^{-1}\,\rm Mpc$ comoving horizontally together with the gas density (from blue to red) within the \hagn simulation at $z=1.2$. As expected, $\mathbf{e}_1$ statistically follows the filaments.}}
\label{fig:image} 
\end{figure*}

The advent of hydrodynamical cosmological simulations is arguably the best way forward to make better predictions on the complex relation between halo shape and spin and galaxy shape and spin. Local studies of the relation between dark matter (DM) and baryonic spins or inertia tensors have been conducted to measure the degree of alignment between the inertia tensor of the DM halo and the stellar component (not weighted by luminosity) \citep{hahn10,Ten++14}. They find a typical excursion of the misalignment angle of $30-10^\circ$ for halos ranging in mass from $10^{10}$ to $10^{14} h^{-1}\msun$.

More recently, \citet{dubois14} measured the alignment of galaxy spins with the large-scale filamentary network in the \hagn simulation, a state-of-the-art hydrodynamical simulation which produced  synthetic galaxies displaying morphological diversity by redshift $z=1.2$. \citeauthor{dubois14} found that simulated galaxies have a spin which is either parallel to their neighbouring filament for low-mass, disc-dominated, blue galaxies, or perpendicular to it for high-mass, velocity dispersion dominated, red galaxies, the rapid reorientation of the latter massive galaxies being due to mergers. This suggests a scenario in which galaxies form in the vorticity-rich neighbourhood of filaments, and then migrate towards the nodes of the cosmic web, converting their orbital momentum into spin.
The inherently anisotropic nature of the large-scale structure (filaments and walls) and its complex imprint on the shape and spin of galaxies { (see also \cite{pichon14})} may prevent isotropic approaches from making accurate predictions and suggest that the imprint of IA might be more severe for higher-order statistics of the cosmic shear signal and definitely not addressable with simple prescriptions for the relation between the spin or inertia tensor of halos and galaxies.

We thus propose here to extend the work of \citet{dubois14} by bringing the  findings of the \hagn simulation (See Fig.~\ref{fig:image}) closer to the framework of weak lensing observables. In particular we  will measure the correlations between galactic spins and their surrounding tidal field (related to the so-called GI term of \citet{H+S04}) and the correlations between spins themselves (related to the so-called II term of the same reference). We will also exhibit the variation of this quantities with the mass and colour of our galaxies, looking for populations where the IA effect is particularly severe or reduced. Our main finding is an excess alignment between the bluest galaxies of our synthetic catalog, and no detectable alignement for the reddest ones. This conclusion is in apparent contradiction with the works cited above, and we will discuss how this can be explained mainly by selection effects (on mass range, redshifts of catalog, etc.).

Throughout this work, we use the stellar spin as a proxy for the ellipticity of galaxies, without attempting to project galactic ellipticities perpendicular to a given sightline. This choice differs from other authors who rather considered the inertia tensor of the stellar mass \citep[see e.g.][]{Ten++14}. We believe spin can give a complementary insight on the apparent luminosity-weighted projected morphology of a generally star-forming galaxy. In addition, as one gets closer to the resolution limit of the simulation, we believe that the reliability of the simulated spin will hold longer than the reliability of the overall shape of the stellar component. Although we 
mainly focus on 3D quantities that are better suited to quantify the physical degree of IA in our simulation, we give some guidelines for inferring projected quantities, in the usual formalism of weak lensing.

The paper is organised as follows. 
Section~\ref{sec:simu} presents the \hagn simulation and describes our method for measuring spin and inertia tensor. It also illustrates how much the shape inferred from spin can be favourably compared to the shape inferred from the inertia tensor of stars, hence supporting our choice of using the spin as a proxy for the ellipticity of galaxies.
Section~\ref{sec:WL} defines how intrinsic alignments are quantified and where they contaminate the weak lensing observables.
In section~\ref{sec:spin-Tij}, we present the cross-correlation between the principle axes of the tidal tensor and the spin vector as a function of distance and further show the zero-lag one-point Probability Distribution Function (PDF) of this angle.
Section~\ref{sec:II}  investigates the spin-spin two-point correlation as a function of separation and the projected ellipticity two-point correlation function.
{ Section~\ref{sec:grid} checks that grid locking effects do not dominate the measurements.}
We finally conclude in Section~\ref{sec:conclusion} and discuss briefly how the statistics depends on the synthetic colours of galaxies. We also sketch how  our findings can be cast into predictions on the contamination of weak lensing by IA.
Appendix~\ref{sec:halos} studies the corresponding alignments for DM halos.

\section{The synthetic universe}
\label{sec:simu}

Let us shortly describe the \hagn simulation (Section~\ref{section:numerics}, see~\citealp{dubois14} for more details) and  explain how galaxy properties are extracted out of it (Section~\ref{section:postprocess}).

\subsection{The Horizon-AGN simulation}
\label{section:numerics}

{
A standard $\Lambda$CDM cosmology compatible with the WMAP-7 cosmology~\citep{komatsuetal11} is adopted, with total matter density $\Omega_{\rm m}=0.272$, dark energy density $\Omega_\Lambda=0.728$, amplitude of the matter power spectrum $\sigma_8=0.81$, baryon density $\Omega_{\rm b}=0.045$, Hubble constant $H_0=70.4 \, \rm km\,s^{-1}\,Mpc^{-1}$, and $n_s=0.967$. The \hagn simulation  has been run with $1024^3$ dark matter (DM) particles
in a $L_{\rm box}=100 \, h^{-1}\rm\,Mpc$ box, so as to obtain a DM mass resolution of $M_{\rm DM, res}=8\times 10^7 \, \rm M_\odot$.
The Adaptive Mesh Refinement code {\sc ramses}~\citep{teyssier02} has been used to run the simulation with
an initial mesh refinement of up to $\Delta x=1\, \rm kpc$ (7 levels of refinement).
The refinement scheme follows a quasi-Lagrangian criterion: if the number of DM particles in a cell is more than 8, or if the total baryonic mass in a cell is 8 times the initial DM mass resolution, a new refinement level is triggered.

A~\cite{sutherland&dopita93} model is used to allow gas cooling by means of H and He cooling down to $10^4\, \rm K$ with a contribution from metals. 
Following~\cite{haardt&madau96}, 
heating from a uniform UV background takes place after redshift $z_{\rm reion} = 10$.
We model metallicity as a passive variable for the gas that varies according to the injection of gas ejecta during supernovae explosions and stellar winds.
A Schmidt law is used to model star formation:
$\dot \rho_*= \epsilon_* {\rho / t_{\rm ff}}\, ,$ where $\dot \rho_*$ is the star formation rate density, $\epsilon_*=0.02$~\citep{kennicutt98, krumholz&tan07} the constant star formation efficiency, and $t_{\rm ff}$ the local free-fall time of the gas.
We allow star formation where
the gas Hydrogen number density exceeds $n_0=0.1\, \rm H\, cm^{-3}$ according to a Poisson random process~\citep{rasera&teyssier06, dubois&teyssier08winds} with a stellar mass resolution of $M_{*, \rm res}=\rho_0 \Delta x^3\simeq 2\times 10^6 \, \rm M_\odot$.

We model
stellar feedback using a \citet{salpeter55} initial mass function with a low-mass (high-mass) cut-off of $0.1\, \rm M_{\odot}$ ($100 \, \rm M_{\odot}$). 
In particular, the mechanical energy from supernovae type II and stellar winds follows the prescription of {\sc starburst99}~\citep{leithereretal99, leithereretal10}, and the frequency of type Ia supernovae explosions is taken from~\cite{greggio&renzini83}. 

\begin{figure*}
\includegraphics[width=0.65\columnwidth]{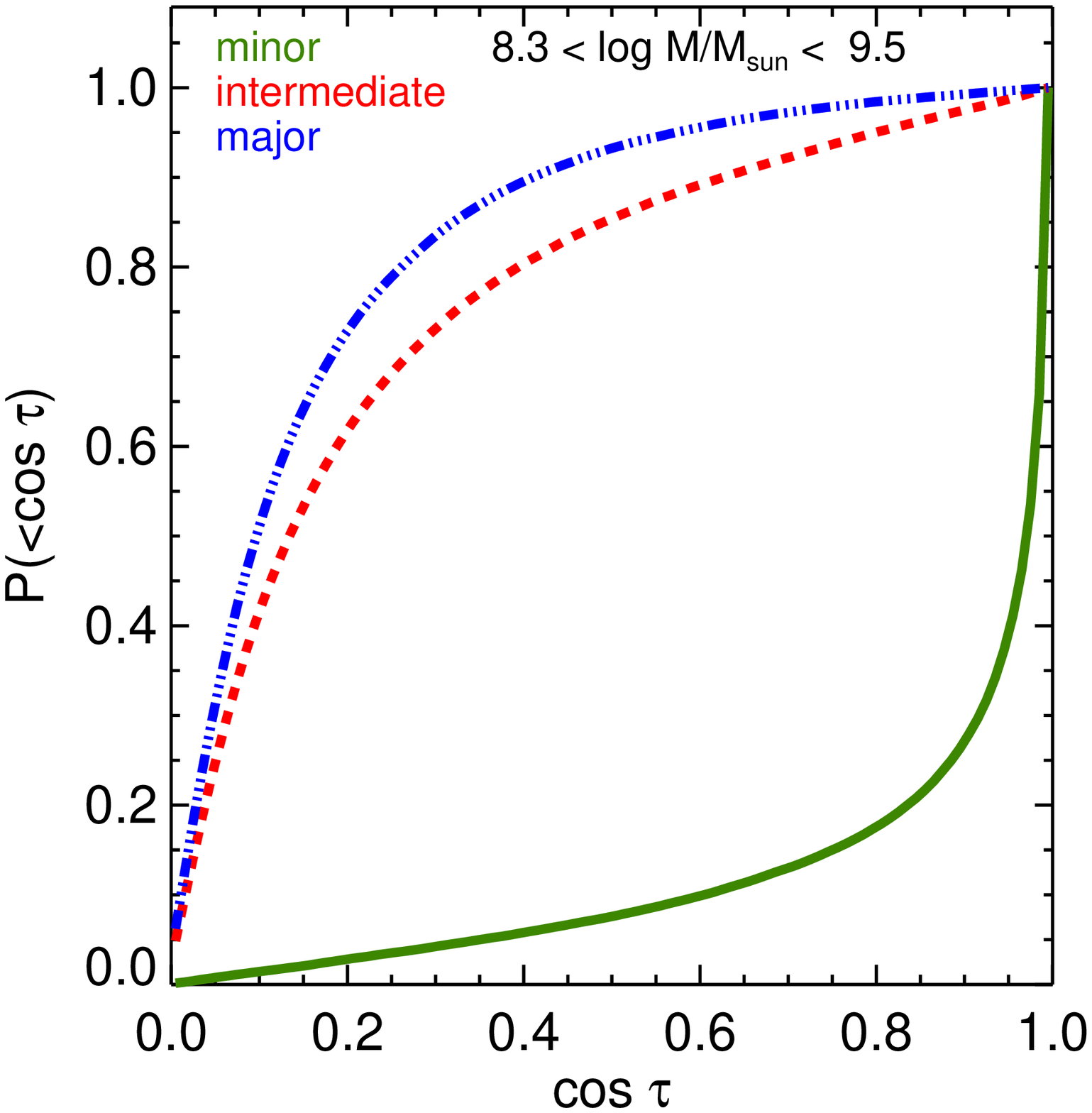}
\includegraphics[width=0.65\columnwidth]{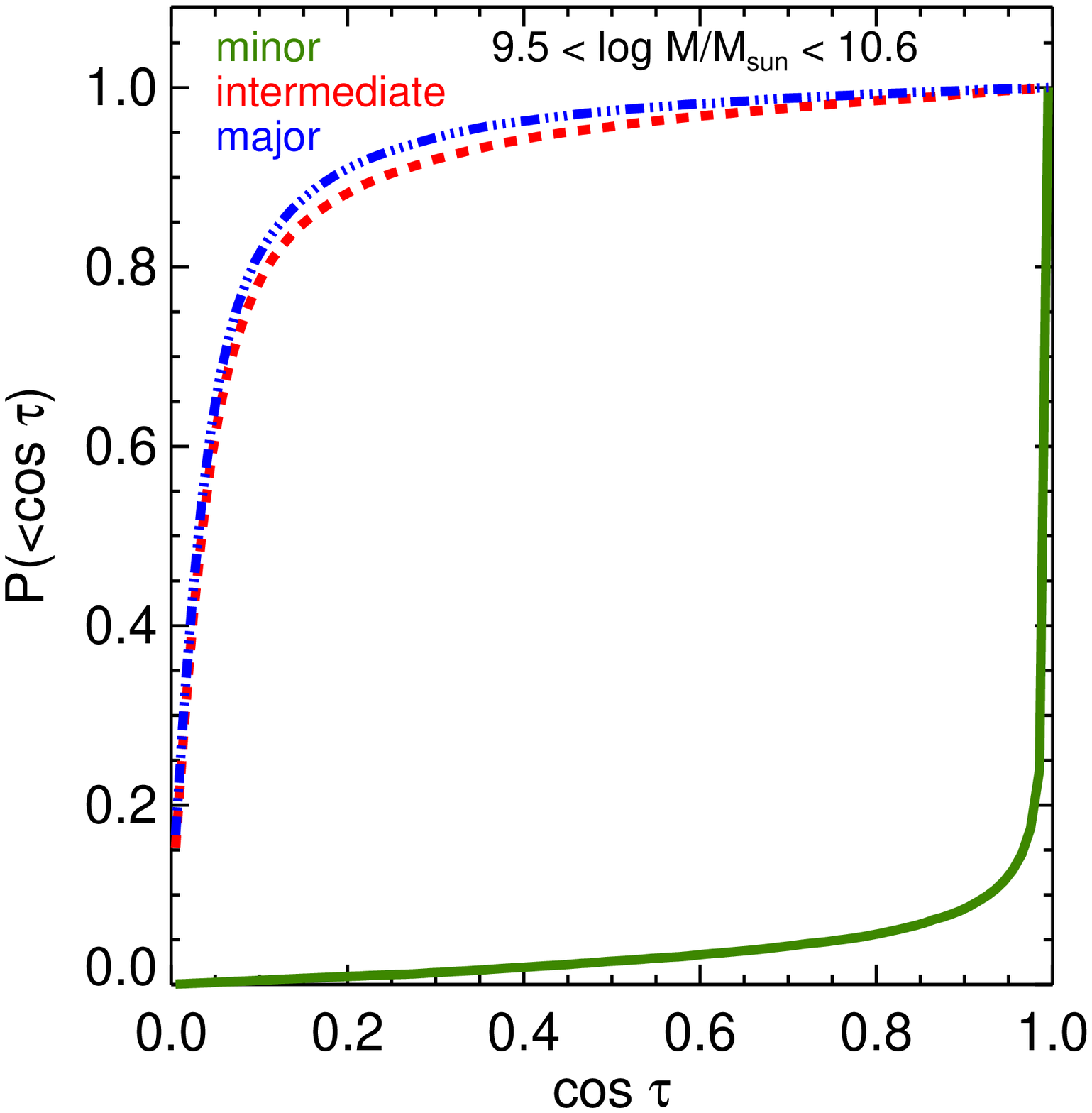}
\includegraphics[width=0.65\columnwidth]{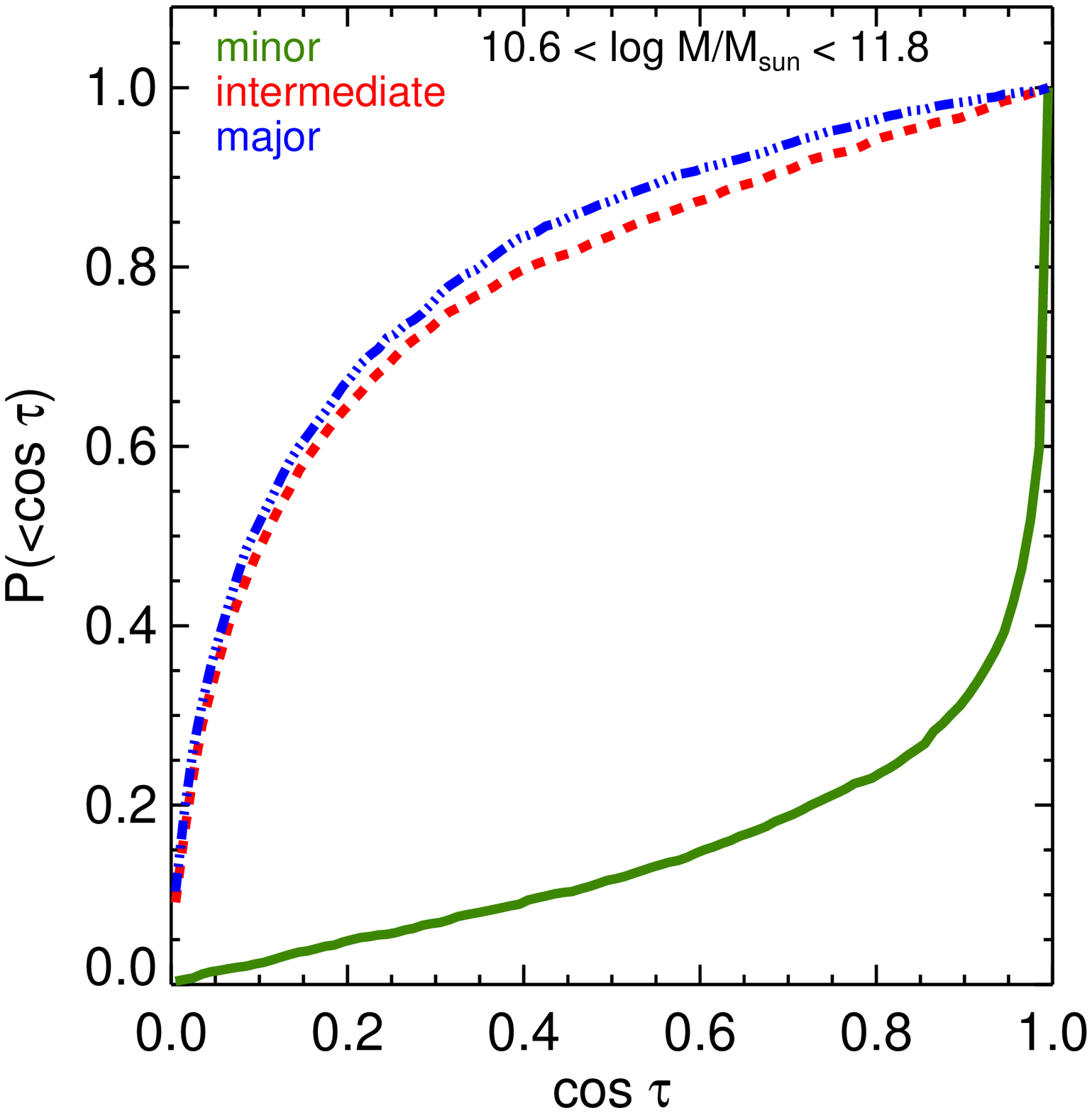}
\caption{Cumulative PDF of the cosine of the angle, $\cos \tau$, between the galaxy spin and the minor (green solid), intermediate (red dashed) and major (blue dot-dashed) axis of the inertia tensor of the galaxy at $z=1.2$. Galaxies tend to align their spin with the minor axis. The most massive galaxies, which are more likely dominated by dispersion rather than rotation, show less correlation.}
\label{fig:spin-ell} 
\end{figure*}

Active Galactic Nuclei (AGN) feedback is modelled according to~\cite{duboisetal12agnmodel}.
A Bondi-Hoyle-Lyttleton accretion rate onto Black Holes is used
 $\dot M_{\rm BH}=4\pi \alpha G^2 M_{\rm BH}^2 \bar \rho / (\bar c_s^2+\bar u^2) ^{3/2},$
where $M_{\rm BH}$ is the BH mass, $\bar \rho$ is the average gas density, $\bar c_s$ is the average sound speed, $\bar u$ is the average gas velocity relative to the BH velocity, and $\alpha$ is a dimensionless boost factor with $\alpha=(\rho/\rho_0)^2$ when $\rho>\rho_0$ and $\alpha=1$ otherwise~\citep{booth&schaye09} in order to account for our inability to capture the colder and higher density regions of the inter-stellar medium.
The effective accretion rate onto BHs is capped at the Eddington accretion rate:
$\dot M_{\rm Edd}=4\pi G M_{\rm BH}m_{\rm p} / (\epsilon_{\rm r} \sigma_{\rm T} c),$
where $\sigma_{\rm T}$ is the Thompson cross-section, $c$ is the speed of light, $m_{\rm p}$ is the proton mass, and $\epsilon_{\rm r}$ is the radiative efficiency, assumed to be equal to $\epsilon_{\rm r}=0.1$ for the \cite{shakura&sunyaev73} accretion onto a Schwarzschild BH. 
Two different modes of
 AGN feedback are accounted for, the \emph{radio} mode operating when $\chi=\dot M_{\rm BH}/\dot M_{\rm Edd}< 0.01$ and the \emph{quasar} mode active otherwise.
 More details  are given in~\citealp{dubois14}.
 }

\subsection{Data analysis}
\label{section:postprocess}

\subsubsection{Galaxy catalogue}

Galaxies are identified with the AdaptaHOP finder~\citep{aubertetal04}, which relies directly on the distribution of star particles to construct the catalogue of galaxies.
20 neighbours are used to compute the local density of each particle.
A local threshold of $\rho_{\rm t}=178$ times the average total matter density is applied to select relevant densities.
{
Note that the galaxy population does not depend sensitively on the exact value chosen for this threshold. Our specific choice reflects the fact that the average density of galaxies located at the centre of galaxy clusters is comparable to that of their host.
}
The force softening (minimum size below which substructures are treated as irrelevant) is of $\sim10$ kpc. 
Only galactic structures identified with more than 50 star particles are included in the mock catalogues. This enables a clear identification of galaxies, including those in the process of merging. 
A galaxy catalogues with  $\sim 165 \, 000$ objects is produced at $z=1.2$ with masses between $1.7\times10^{8}$ and $1.4\times10^{12}\, \rm M_{\odot}$.
The galaxy stellar masses quoted in this paper should be understood as the sum over all star particles that belong to a galaxy structure identified by AdaptaHOP. 
{ Note that most results are derived from a subsample of galaxies with a mass above $10^{9}M_{\odot}$, which corresponds to 300 stellar particles.}

\subsubsection{Spin and shape of galaxies}\label{ssseq:spin-shape}
To assign a spin to the galaxies, we compute the total angular momentum of the star particles which make up a given galactic structure relative to the particle of maximum density (centre of the galaxy). To identify the latter, we use the smoothed stellar density constructed with the AdaptaHOP algorithm.
We can therefore write the intrinsic angular momentum vector $\mathbf{L}$ or spin of a galaxy as
\begin{equation}
\label{eq:spindef}
  \mathbf{L} = \sum_{ \alpha=1}^{N} m^{( \alpha)} \mathbf{x}^{(\alpha)} \times \mathbf{v}^{(\alpha)} \, ,
\end{equation}
where the $\alpha$ upperscript denotes the $\alpha$-th stellar particle
of mass $m^{ (\alpha)}$, 
position
$\mathbf{x}^{ (\alpha)}$ and velocity
  $\mathbf{v}^{(\alpha)}$ relative to the center of mass of that galaxy. 
Likewise, we also measure the (reduced) inertia tensor of a galaxy:
\begin{equation}\label{eq:inerdef}
I_{ ij}={\displaystyle\sum_{ \alpha=1}^{N} m^{( \alpha)} x^{(\alpha)}_{i} x^{(\alpha)}_{j} \over \displaystyle \sum_{\alpha=1}^{N} m^{ (\alpha)}}\, .
\end{equation}
This inertia tensor is then diagonalised to obtain the eigenvalues $\lambda_1\le \lambda_2 \le \lambda_3$ and the corresponding unit eigenvectors $\mathbf{u}_1$, $\mathbf{u}_2$ and $\mathbf{u}_3$ (respectively minor, intermediate and major axis of the ellipsoid).

Fig.~\ref{fig:spin-ell} shows the PDF of the angle $\tau_{i}$ between the spin and each of the principle axes $\mathbf{u}_i$ of the galaxies:
\begin{equation}\label{eq:taudef}
\cos \tau_{ i} = \frac{\mathbf{L} . \mathbf{u}_{i} }{\vert \mathbf{L} \vert}\,.
\end{equation}
As expected, galaxies tend to have a spin well-aligned with the minor axis ($\mathbf{u}_1$) of the inertia tensor, with a mean value $\langle \cos \tau_1 \rangle = 0.90$. 
The correlation is slightly less pronounced for the most massive galaxies, for which the rotation support is weaker compared to velocity dispersion, but it still shows a strong degree of alignment ($\langle \cos \tau_1 \rangle = 0.85$).

Owing to the tight alignment between the spin of galaxies and the minor axis of the inertia tensor, we expect that an analysis of galactic spin or inertia tensor orientations will capture the physical mechanisms producing IA equally well. For weak lensing predictions however, we shall pay attention to the modulus of the ellipticity and not only its direction when turning spins into projected ellipticities. 
Namely, the projected ellipticity of an axisymmetric disky galaxy depends on the disk thickness \citep[e.g.][]{Joa++13a,Joa++13b}. If we note $q_d$, the disk flattening or axis ratio, the apparent axis ratio $q_p$ of a projected galaxy along the line of sight aligned with the $z$ axis, reads:
\begin{equation}\label{eq:L2q}
  q_p = \frac{\vert L_z\vert}{\vert \mathbf{L} \vert} + q_d \sqrt{ 1 - \frac{L_z^2}{\vert \mathbf{L}\vert^2}}\;.
\end{equation}
In principle, we could measure the flattening of the simulated galaxies to infer the projected axis ratios. 
Unfortunately, the finite $1\, h^{-1}\, \rm kpc$ resolution of the {\hagn}simulation overestimates the thickness of the disk of low-mass galaxies. 
To alleviate this problem, we will assume $q_d=0$ in the remainder of this work, and thus maximise the moduli of the projected ellipticities. This can be seen as a conservative approach since we effectively maximise the implication of spin IA on apparent alignments of projected ellipticities, either for the correlation between spins themselves (II) of between spins and tidal tensor (GI). We shall come back to the fidelity of our simulations at recovering the eigenvalues of the inertia tensor and not only its eigen-directions in a future work.

\subsubsection{Rest-frame intrinsic colours}
\label{sec:colours}
In order to ascertain the sensitivity of our measurements to galaxy colours, we compute the absolute AB magnitudes and rest-frame colours of the mock galaxies using single stellar population models from~\cite{bruzual&charlot03} adopting a Salpeter initial mass function. 
Each star particle contributes a flux per frequency that depends on its mass, age and metallicity. 
The sum of the contribution of all star particles is passed through u, g, r, or i filter bands from the SDSS. 
Fluxes are expressed as rest-frame quantities (i.e. that do not take into account the red-shifting of spectra) and, for the sake of simplicity, dust extinction is neglected.
Once all the star particles have been assigned a flux in each of the colour channels, we build the 2D projected maps for individual galaxies (satellites are excised with the galaxy finder). Summing up the contribution of their stars yields the galaxy luminosity in a given filter band.

Thorough this work, we will investigate the alignment properties of galaxies filtered by their colours. We separate our catalogue in three colour bins (in $u-r$) such that the number of galaxy in each is identical. Our bluest subset correspond to  $u-r<0.78$ while the reddest one have $u-r>1.1$. This choice insures a similar noise level in all of our measurements.

\begin{figure}
\includegraphics[width=0.9\columnwidth]{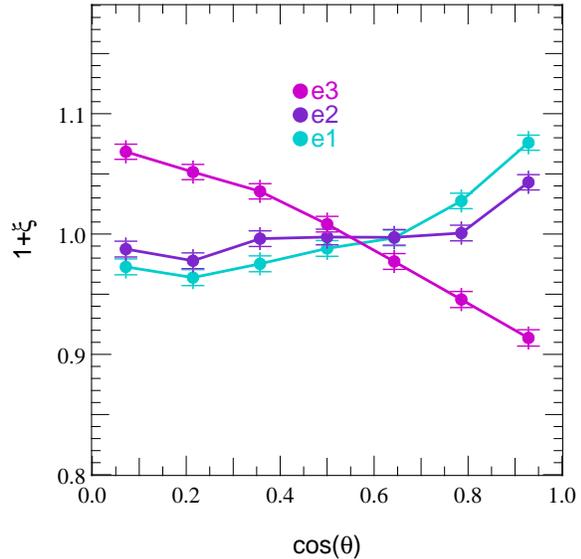}
\caption{Excess probability of alignment between the spin of galaxies and the minor, intermediate, or major axis (resp. cyan, purple and magenta) 
of the tidal tensor in the Horizon-AGN simulation.
The  error bars represent the Poisson noise.
The spin of galaxies tends to align with the minor eigen-direction. Fig~\ref{fig:tidaltensor-mass} and Fig.~\ref{fig:tidaltensor-colors} investigate how this alignment changes with respectively galactic mass and colour. 
\label{fig:tidaltensor-gal} 
}
\end{figure}

\section{Intrinsic alignments in the context of cosmic shear studies}
\label{sec:WL}
Weak lensing uses the apparent deformation of the shapes of galaxies on the sky to map the gravitational potential or measure its statistical properties like its power spectrum. The projected surface mass density, integrated along the line of sight to distant sources is often preferred over the potential although they are trivially related by the Poisson equation. The effective convergence $\kappa$, which is nothing but the dimensionless projected density, is statistically described by its power-spectrum $P_\kappa(\ell)$ as a function of wavenumber $\ell$.
For a source at comoving distance $\chi_s$, we can write the convergence at an angular position $\mathbf{\theta}$ \citep{BS01}\footnote{Simplified to a flat Universe case}:
\begin{eqnarray}
 \kappa(\mathbf{\theta},\chi_s) &= &\frac{1}{c^2} \int_0^{\chi_s} \der\chi\, \frac{(\chi_s-\chi)\chi}{\chi_s} \left[ \frac{\partial^2}{\partial x^2} + \frac{\partial^2}{\partial y^2 }\right] \Phi, \nonumber\\
 &=& \frac{3 H_0^2 \Omega_0 }{2 c^2} \int_0^{\chi_s} \der\chi\, \frac{(\chi_s-\chi)\chi}{\chi_s} \frac{\delta( \chi \mathbf{\theta}, \chi)}{a(\chi)},
\end{eqnarray}
where $\chi$ is the comoving distance, $a$ the expansion factor, $\delta$, the density contrast, and $\Phi$, the three-dimensional gravitational potential, are related by the Poisson equation:
\begin{equation}\label{eq:poisson}
\Delta \Phi = \frac{3 H_0^2 \Omega_0}{2 a} \delta\,.
\end{equation}
This can easily be generalised to a population of sources with a broad redshift distribution \citep{BS01}.

The relation between $\Phi$ and $\delta$ can be cast into a relation between the lensing potential $\phi$ and effective convergence $\kappa$,
and the effective shear $\gamma_i$, which involves the traceless parts  of the projected tidal tensor.  All these quantities are defined by:
\begin{align}
\phi &= \displaystyle\frac{2}{c^2}  \int_0^{\chi_s} \der\chi\, \frac{(\chi_s-\chi)\chi}{\chi_s} \Phi \,, \,\,\, \displaystyle
\kappa \ = \frac{1}{2} \left( \frac{\partial^2 \phi}{\partial \theta_1^2} + \frac{\partial^2 \phi}{\partial \theta_2^2} \right)\!,\\
\gamma_1  &=  \frac{1}{2} \left( \frac{\partial^2 \phi}{\partial \theta_1^2} - \frac{\partial^2 \phi}{\partial \theta_2^2} \right) \, , \quad
\gamma_2  =  \frac{\partial^2 \phi}{\partial \theta_1 \partial \theta_2}\;.
\end{align}
It is generally suitable to treat the shear in complex notations $\gamma = \gamma_1 + i \gamma_2 $.
This quantity is most easily accessible as it captures the amount of anisotropic distortion a light bundle experiences on its way from a distant source to the observer. Therefore, the observed ellipticity of such a source, in the weak lensing regime of small distortions, is directly related to the shear. Indeed, by also defining a complex ellipticity $e=e_1 + i e_2 = |e| e^{2 i\psi}$, such that $|e|=({1-q})/({1+q})$ and $q=b/a$ is the major ($a$) to minor ($b$) axis ratio, we have 
\begin{equation}\label{eq:e2g}
  e = e_s + \gamma\;,
\end{equation}
where $e$ is the apparent ellipticity and $e_s$ the intrinsic source ellipticity (the one we would have observed without lensing).

An important statistics of this cosmic shear distortion field is the two-point correlation of projected ellipticities that can formally be split into the following components:
\begin{equation}
\label{eq:ee}
\left\langle e(\vartheta) e(\vartheta+\theta) \right\rangle_\vartheta = \left\langle e_{s}e_{s}' \right\rangle+2\left\langle e_{s} \gamma'\right\rangle
+\left\langle \gamma \gamma' \right\rangle\,,
\end{equation}
where, for compactness, the prime means at an angular distance $\theta$ from the first location.
The cosmological weak lensing signal is commonly decomposed into the $\xi_+$ and $\xi_-$ shear correlation functions. Following \cite{Sch++02}, $\xi_{\pm}$ is given by
\begin{equation}
\xi_{\pm}(\theta) = \langle \gamma_+ \gamma_+ \rangle  \pm  \langle \gamma_\times \gamma_\times\rangle 
= \frac{1}{2\pi} \int_0^\infty \der\ell\, \ell P_\kappa(\ell) J_{0/4}(\ell \theta)\;, \nonumber
\end{equation}
where $J_0$ and $J_4$ are the 0-th and 4-th order Bessel functions for $\xi_+$ and $\xi_-$ respectively. In this expression, $\gamma_+$ (resp. $\gamma_{\times}$) is the component of the complex shear orientated 0/90$^o$ (resp. $\pm$45$^o$) with respect to the line connecting two galaxies separated by a projected distance $\theta$. 

The fundamental assumption of weak lensing, which allows to infer shear properties from observed ellipticities is that, on average, the intrinsic orientation of sources is completely random. The breakdown of this hypothesis yields additional terms to $\left\langle \gamma \gamma' \right\rangle$ on the right-hand side of Eq.~\eqref{eq:ee} that have to be carefully accounted for in observations. The weak lensing signal is therefore contaminated by the two kinds of IA:
\begin{itemize}
\item the so-called ``II''  term $\left\langle e_{s}e_{s}' \right\rangle$ induced by the intrinsic correlation of the shape of galaxies in the source plane \citep{HRH00,C+M00,Cat++01}. This mostly concerns pairs of galaxies that are at similar redshifts.
\item and the so-called ``GI'' term $\left\langle e_{s} \gamma'\right\rangle$ coming from correlation between the intrinsic ellipticity of a galaxy and the induced ellipticity (or shear) of a source at higher redshift \citep{H+S04,Hey++06,Joa++11}. This non-trivial term is indirectly explained if the shape of galaxies is correlated with the local gravitational tidal field, which also contributes to the shear signal experienced by the far source in a given pair of observed ellipticities.
\end{itemize}
In this work we propose to measure these two effects in the \hagn simulation, Section~\ref{sec:spin-tij}
 being devoted to the ``GI'' term (essentially captured by spin-tidal field correlations) and Section~\ref{sec:II} to the ``II'' term (essentially captured by spin-spin correlations). 

Before presenting those results, we also give here some guidelines on the way projected correlation functions are worked out in the simulation. In Section~\ref{ssseq:spin-shape}, we presented our method for relating three-dimensional galaxy spins $\mathbf{L}$ to projected ellipticities in the plane of the sky. It is based on the ansatz that the axis ratio of our galaxies is well approximated by $q= {\vert L_z\vert}/{\vert \mathbf{L}\vert}$, where $z$ is the line of sight direction. The orientation of the major axis of the projected ellipse is $\psi=\pi/2-\arctan (L_y/L_x)$. The projected ellipticities can easily be mapped from cartesian (1,2) coordinates to the $(+,\times)$ frame attached to the separation of a given galaxy pair according to the geometric transformation
\begin{eqnarray}
e_+ &=& -e_1 \cos(2\beta) - e_2 \sin(2\beta)\,, \\
e_\times &=& \phantom{+}e_1 \sin(2\beta) - e_2 \cos(2\beta)\;,
\end{eqnarray}
where $\beta$ is the angle between the separation and the first cartesian coordinate.

With those prescriptions, we can estimate the projected correlation functions for a given projected separation $\theta$. For the II component (dropping the subscript ${s}$), this reads
\begin{equation}
  \xi_{+}^{\rm II}(\theta)=\left\langle e_{+}e'_{+}+e_{\times}e'_{\times}\right\rangle\,.
\end{equation}
Beyond this 2D measurement, and to limit the dilution of the IA signal with projected angular distances, we will also measure the correlation as a function of the 3D comoving galaxy separation whilst still considering 2D ellipticities as the result of the projection along a specific line of sight. We call this correlation function $\eta(r)$ following the notations of \citet{Hey++06} and \citet{Joa++13b}
\begin{equation}
  \eta(r) = \langle e_+(\mathbf{x}) e_+(\mathbf{x}+\mathbf{r}) +  e_\times(\mathbf{x}) e_\times(\mathbf{x}+\mathbf{r}) \rangle_x \;,
\end{equation}
where $\mathbf{r}$ is the 3D galaxy pair separation.

\begin{figure*}
\includegraphics[width=0.65\columnwidth]{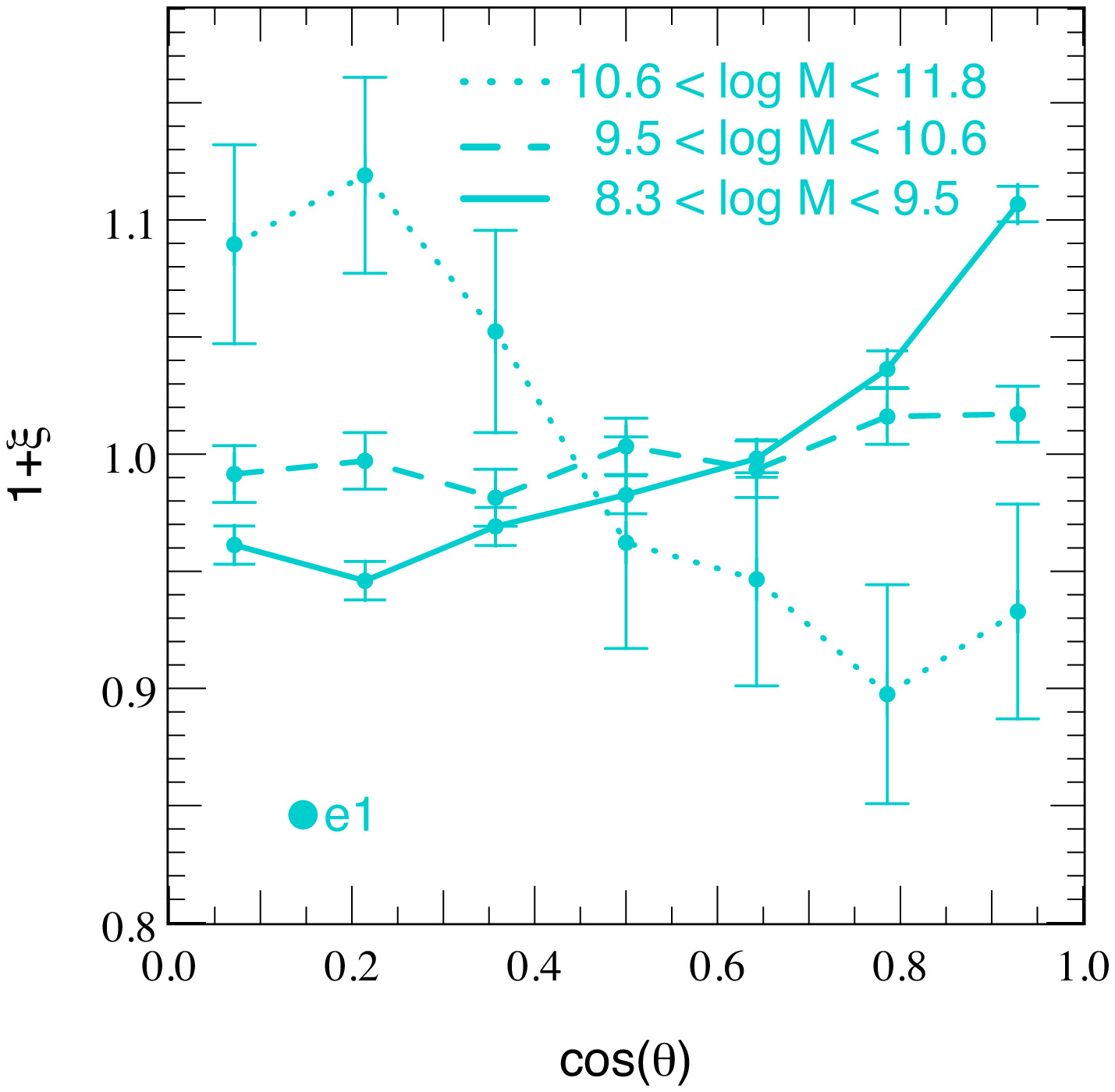}
\includegraphics[width=0.65\columnwidth]{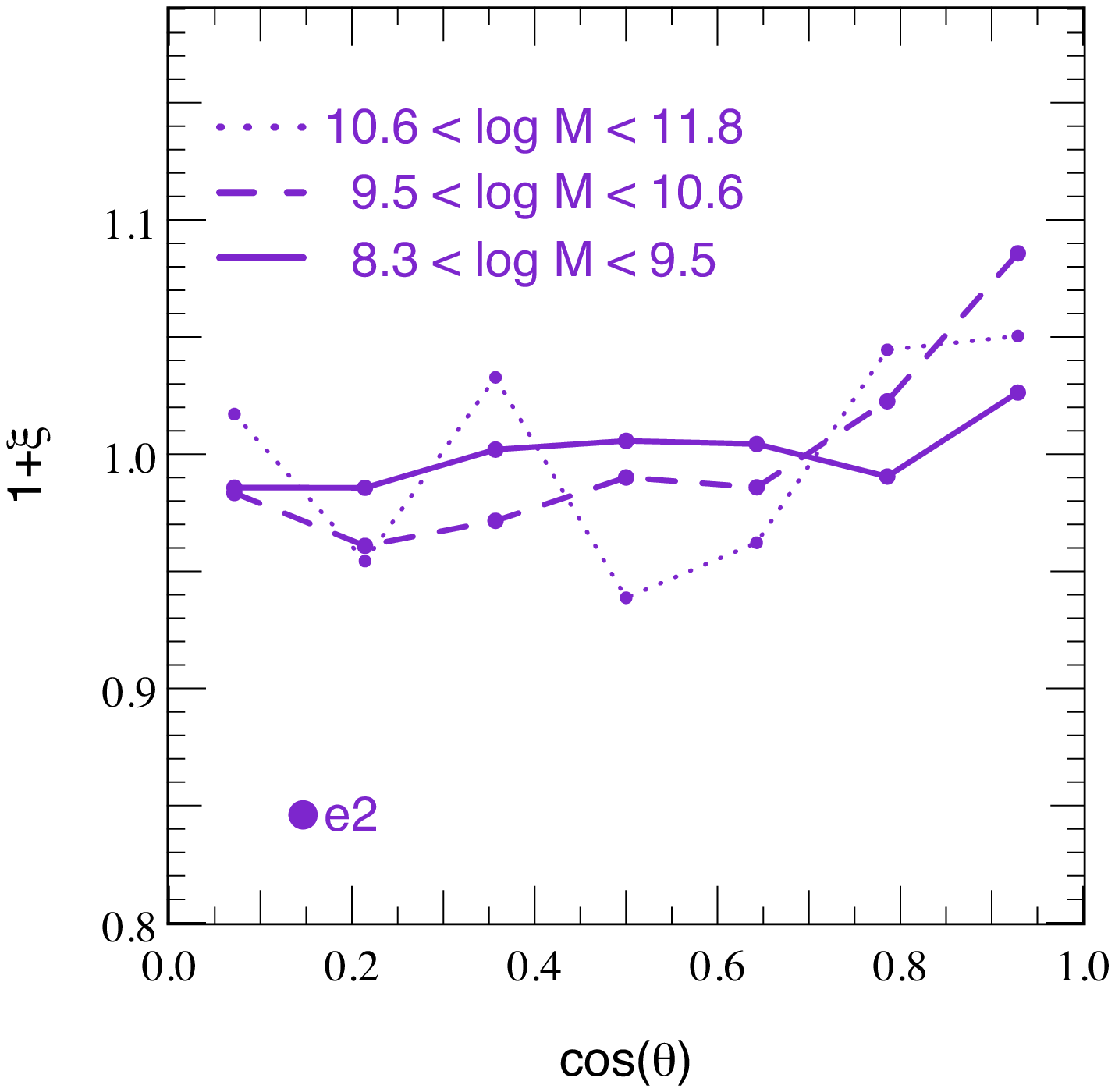}
\includegraphics[width=0.65\columnwidth]{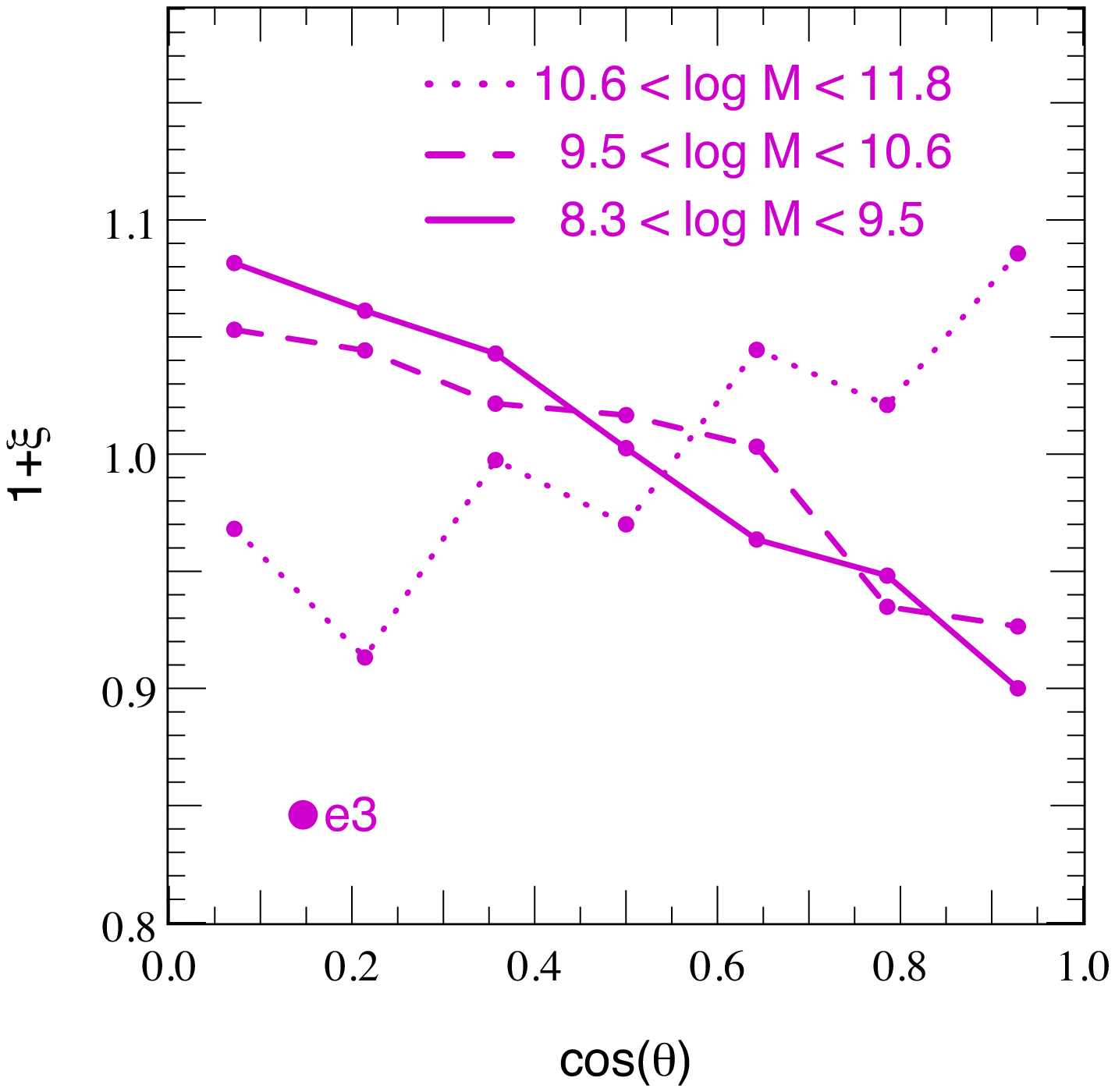}
\caption{PDF of the cosine of the angle between the spin of galaxies and the minor/intermediate/major axis ({from left to right}) 
of the tidal tensor in the Horizon-AGN simulation
when the sample is separated into three different mass bins (solid lines for stellar mass between $2\times  10^{8}$ and $3\times10^{9}\,\rm M_\odot$, dashed lines for stellar mass between $3\times  10^{9}$ and $4\times10^{10} \, \rm M_\odot$ and dotted lines  for stellar mass between $4\times  10^{10}$ and $6\times10^{11}\, \rm M_\odot$). 
The 
error bars represent the Poisson noise and are only shown for $\mathbf{e}_{1}$ (left panel) since they are the same for $\mathbf{e}_{2}$ (middle panel) and $\mathbf{e}_{3}$ (right panel).
The spin of galaxies tends to align with the minor eigen-direction at small mass
and becomes perpendicular to it at larger mass.
\label{fig:tidaltensor-mass} }
\end{figure*}

\begin{figure*}
\includegraphics[width=0.65\columnwidth]{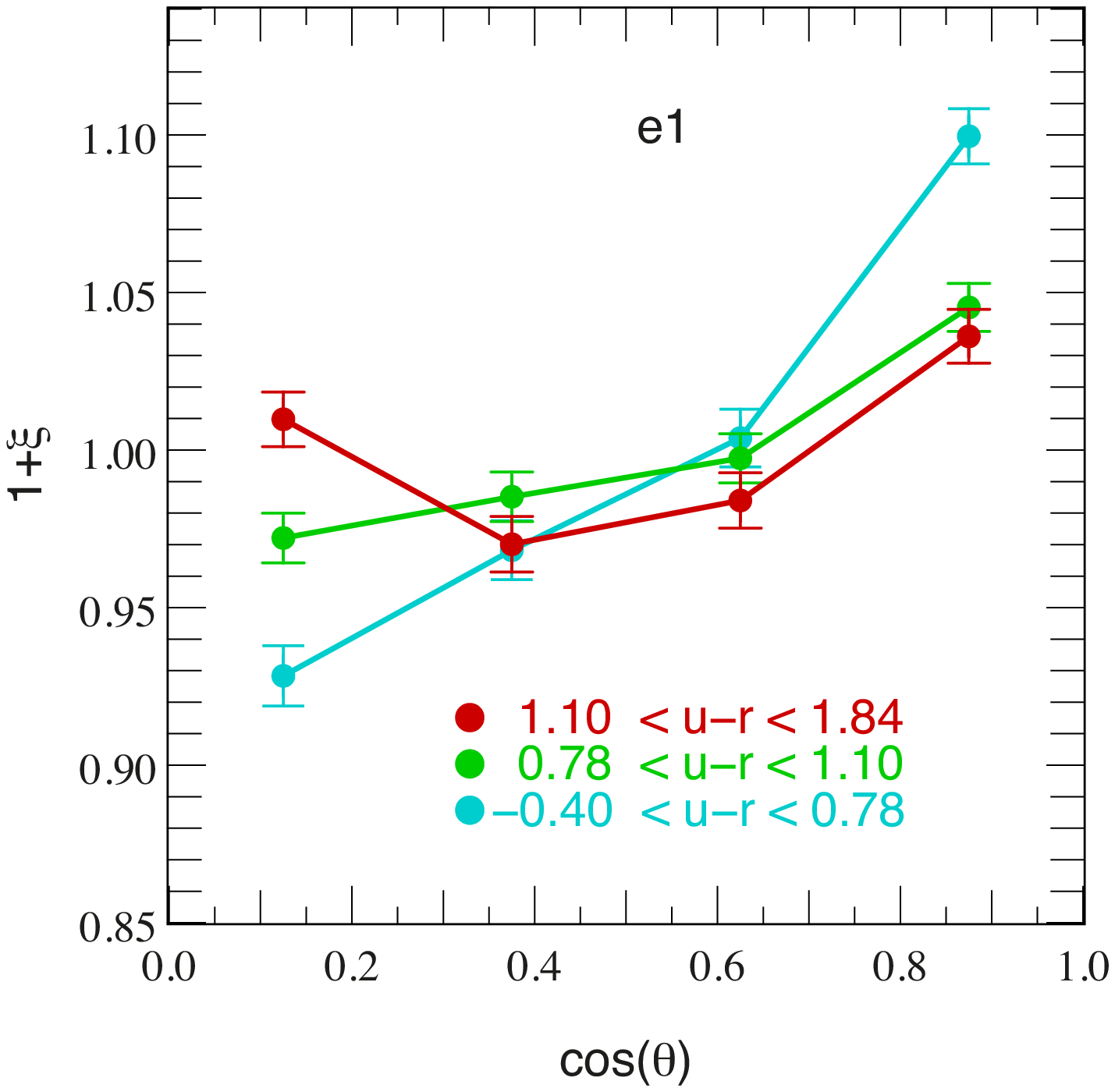}
\includegraphics[width=0.65\columnwidth]{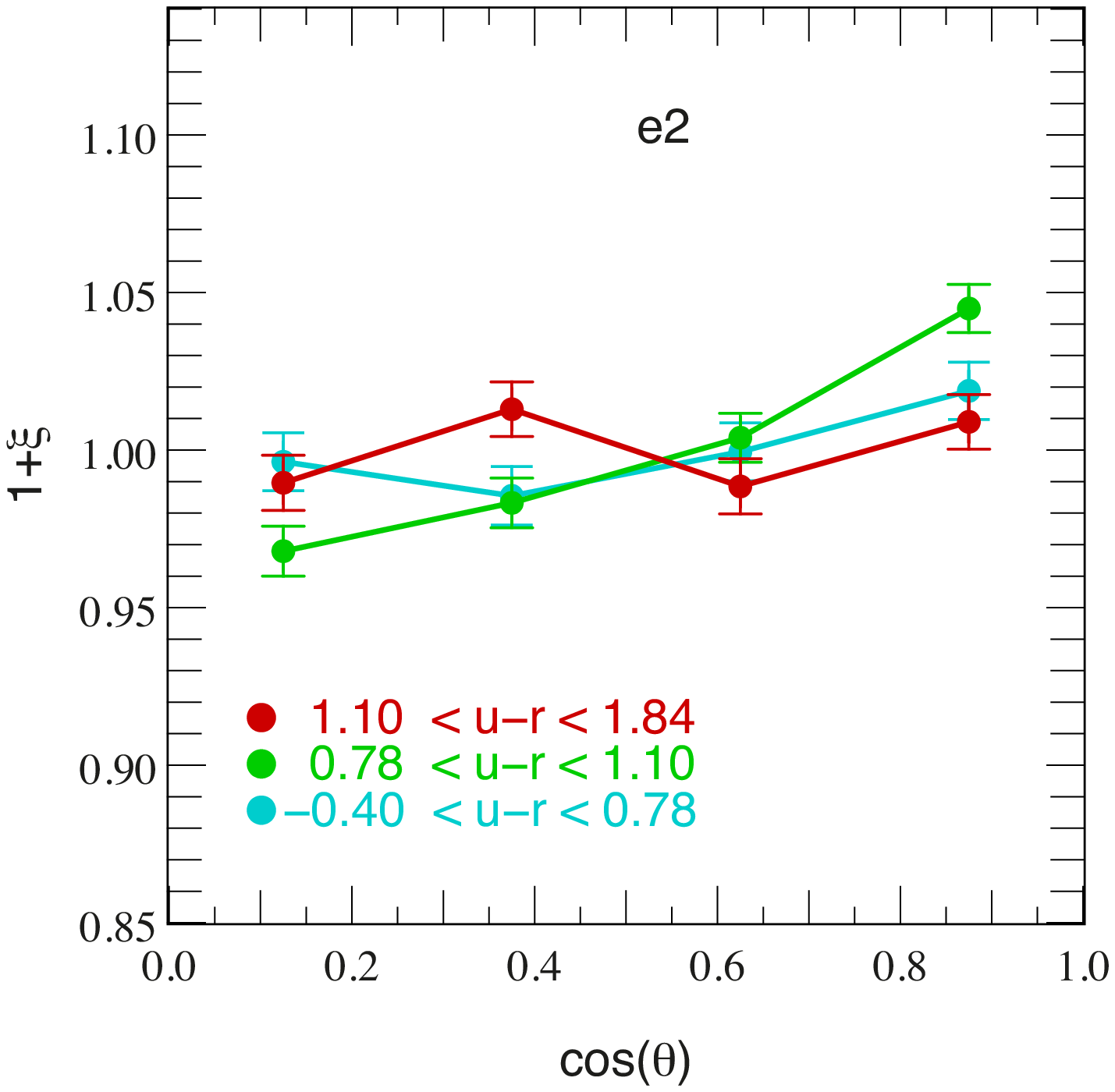}
\includegraphics[width=0.65 \columnwidth]{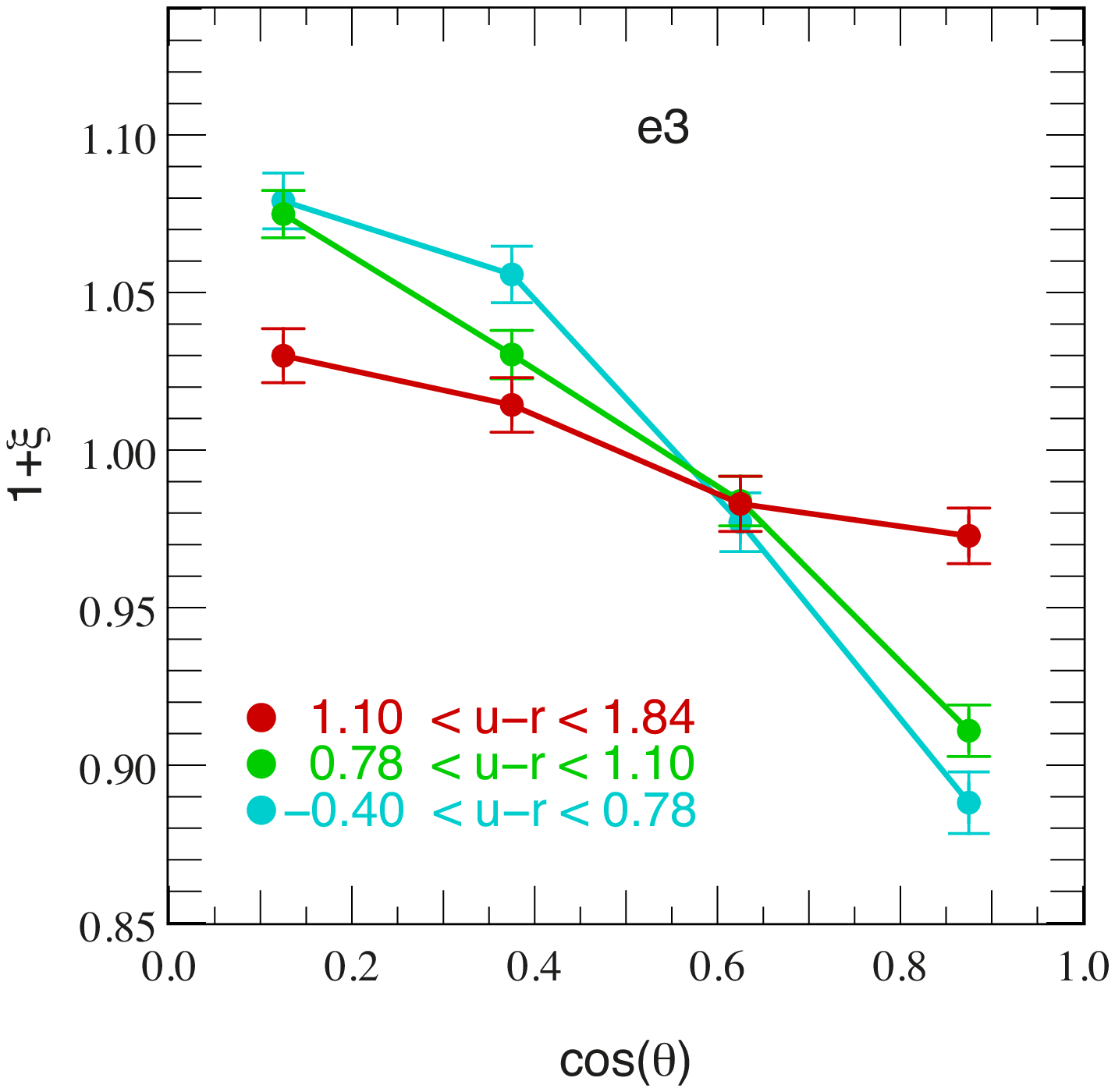}
\caption{Same as Fig.~\ref{fig:tidaltensor-mass} but for different galaxy colours as labeled, meaning that  the left, middle and right panels respectively show the PDF of the angle between  the  $\mathbf{e}_{1}$, $\mathbf{e}_{2}$, $\mathbf{e}_{3}$ directions of the tidal tensor and the galactic spins.
The bluer the galaxy the larger the correlations with the surrounding tidal field. Hence red galaxies are less sensitive to IA.
\label{fig:tidaltensor-colors} 
}
\end{figure*}

\section{Spin-tidal tensor correlations}
\label{sec:spin-tij}
\label{sec:spin-Tij}
In the context of weak lensing surveys, IA can occur through correlations between the shear induced by the gravitational potential in the lens plane, and the intrinsic ellipticity of galaxies in the source plane. Here, we aim to assess the extent to which the tidal tensor and the galactic spins correlate in the  \hagn simulation described in Section~\ref{sec:simu}. For this purpose, we first measure in Section~\ref{sec:onept-spin-tidal} the one-point correlation between the spins and the tidal tensor. Then, Section~\ref{sec:twopt-spin-tidal} is devoted to the measurements of the two-point correlations between the spins and the tidal tensor as a function of the separation.

To study the correlations between the spin direction and the surrounding gravitational tidal field of the galaxies resolved in the \hagn simulation, we measure the components of the 3D (traceless) tidal shear tensor defined as
\begin{equation}
  T_{ij} = \partial_{ij} \Phi - \frac{1}{3} \Delta \Phi \,{\delta}_{ij}\,,
\end{equation}
where $\Phi$ is the gravitational potential and $\delta_{ij}$ the Kronecker delta function. 
The minor, intermediate and major eigen-directions of the tidal tensor $T_{ij}$ are called
$\mathbf{e}_1$, $\mathbf{e}_2$ and $\mathbf{e}_3$ corresponding to the ordered eigenvalues 
$\lambda_{1}\le\lambda_{2}\le\lambda_{3}$ 
of the Hessian of the gravitational potential,
$\partial_{ij} \Phi$
(with which the tidal tensor shares 
the eigen-directions).
In the filamentary regions, 
$\mathbf{e}_1$ gives the direction of the filament (see Fig.~\ref{fig:image}), while the walls
are collapsing along $\mathbf{e}_3$ and extend, locally, in the 
plane spanned by $\mathbf{e}_1$ and $\mathbf{e}_2$ \citep{Pogosyanetal1998}. 

The tidal shear tensor smoothed on scale $R_{s}$, $T_{ij}=\partial_{ij} \Phi_{R_{s}}-\Delta \Phi_{R_{s}}\,{\delta}_{ij}/3$, is computed via Fast Fourier Transform of the density field (including dark matter, stars, gas and black holes) sampled on a $512^{3}$ cartesian grid and convolved with a Gaussian filter of comoving scale $R_{s}=200\, h^{-1}Ê\, \rm  kpc$
\begin{equation}
\partial_{ij} \Phi_{R_{s}}(\mathbf{x})=\frac{3 H_0^2 \Omega_0}{2 a}\int \mathrm{d}^{3}\mathbf{k}\;\delta(\mathbf{k}) \frac{k_{i}k_{j}}{k^{2}}W_{G}(k R_{s})\exp\left({i \,\mathbf{k}\!\cdot\!\mathbf{x}}\right)\,, \nonumber
\end{equation}
where $\delta(\mathbf{k})$ is the Fourier transform of the sampled density field and $W_{G}$ a Gaussian filter.

\begin{figure*}
\includegraphics[width=0.65\columnwidth]{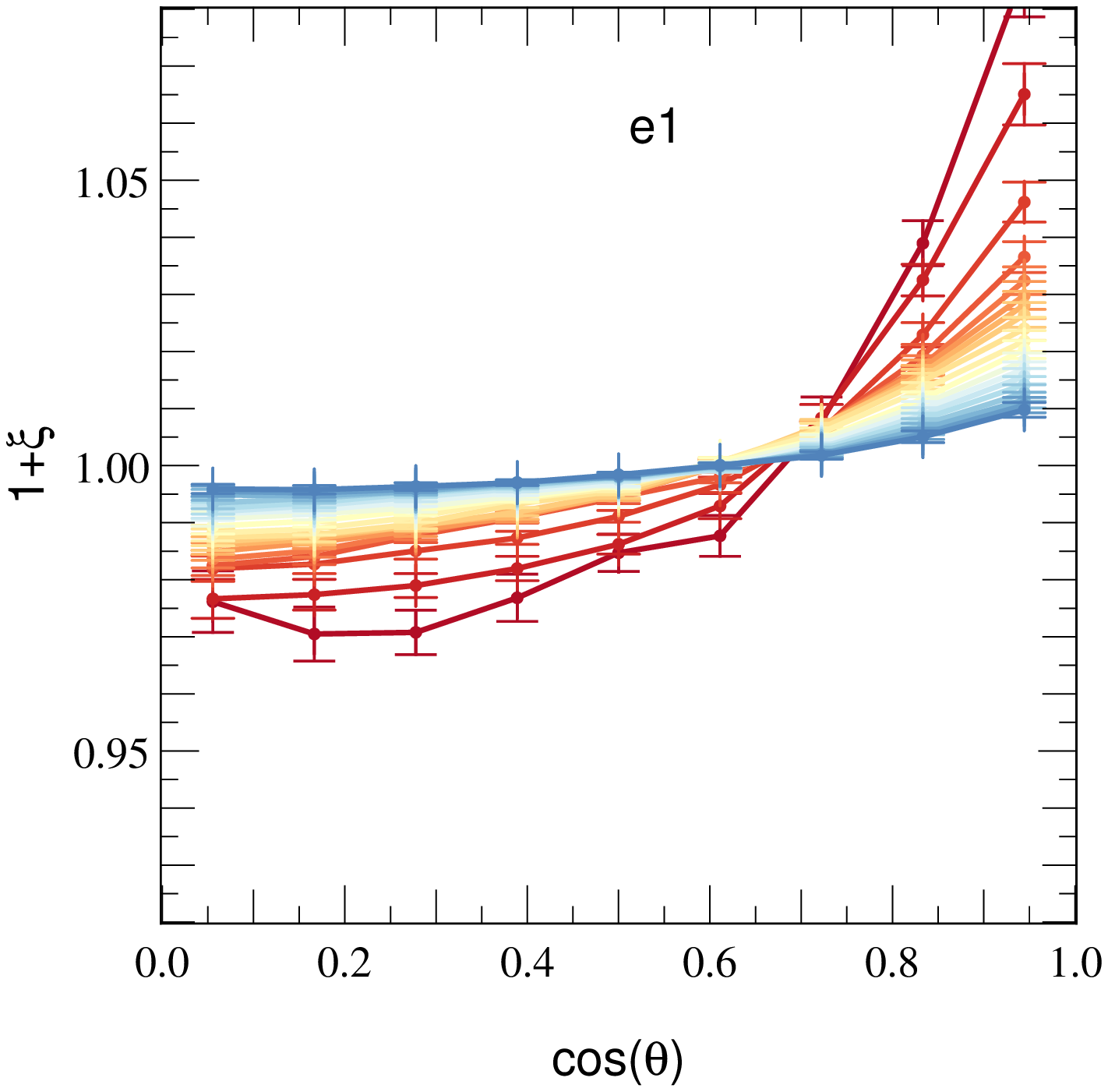}
\includegraphics[width=0.65\columnwidth]{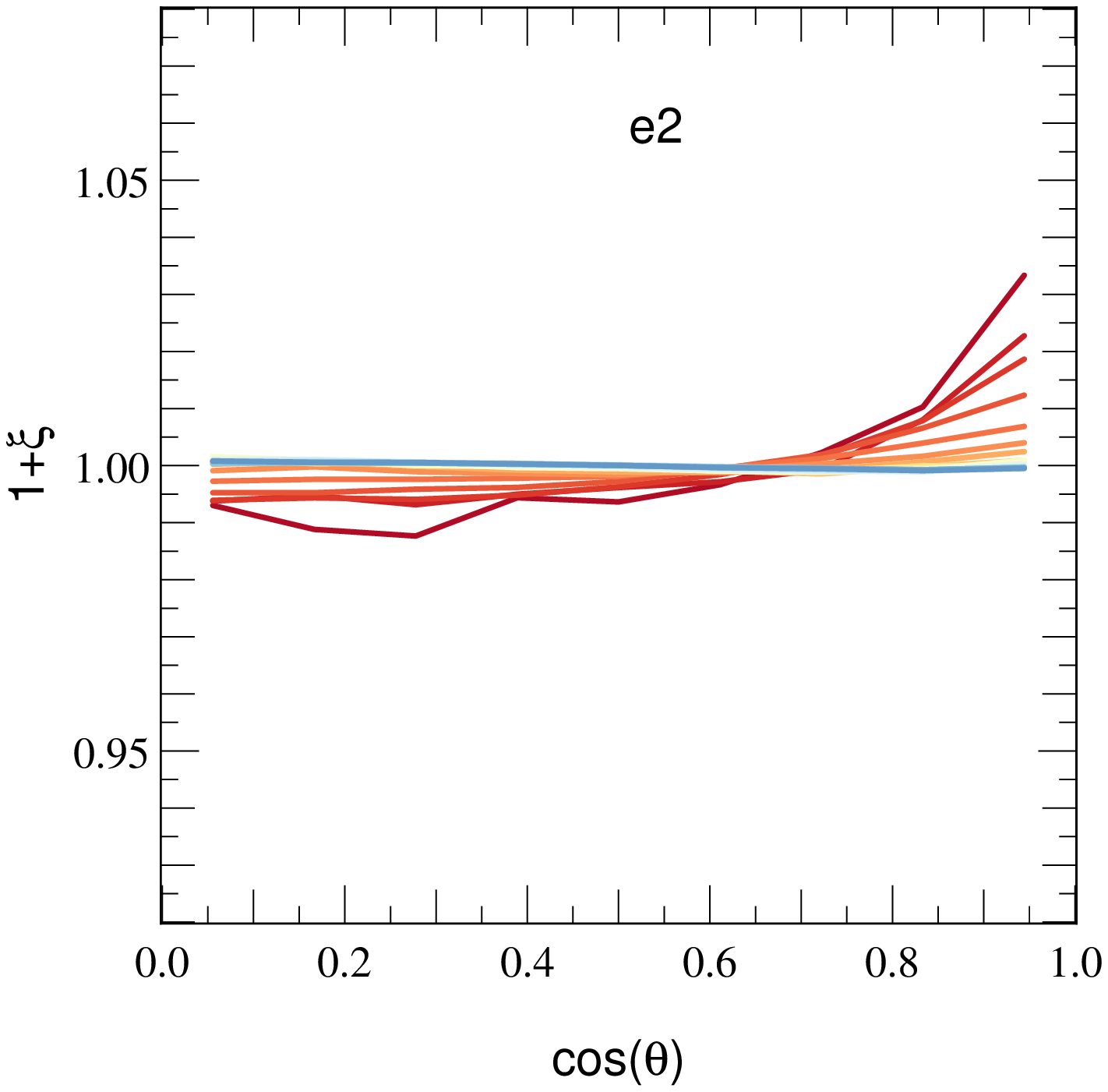}
\includegraphics[width=0.75\columnwidth]{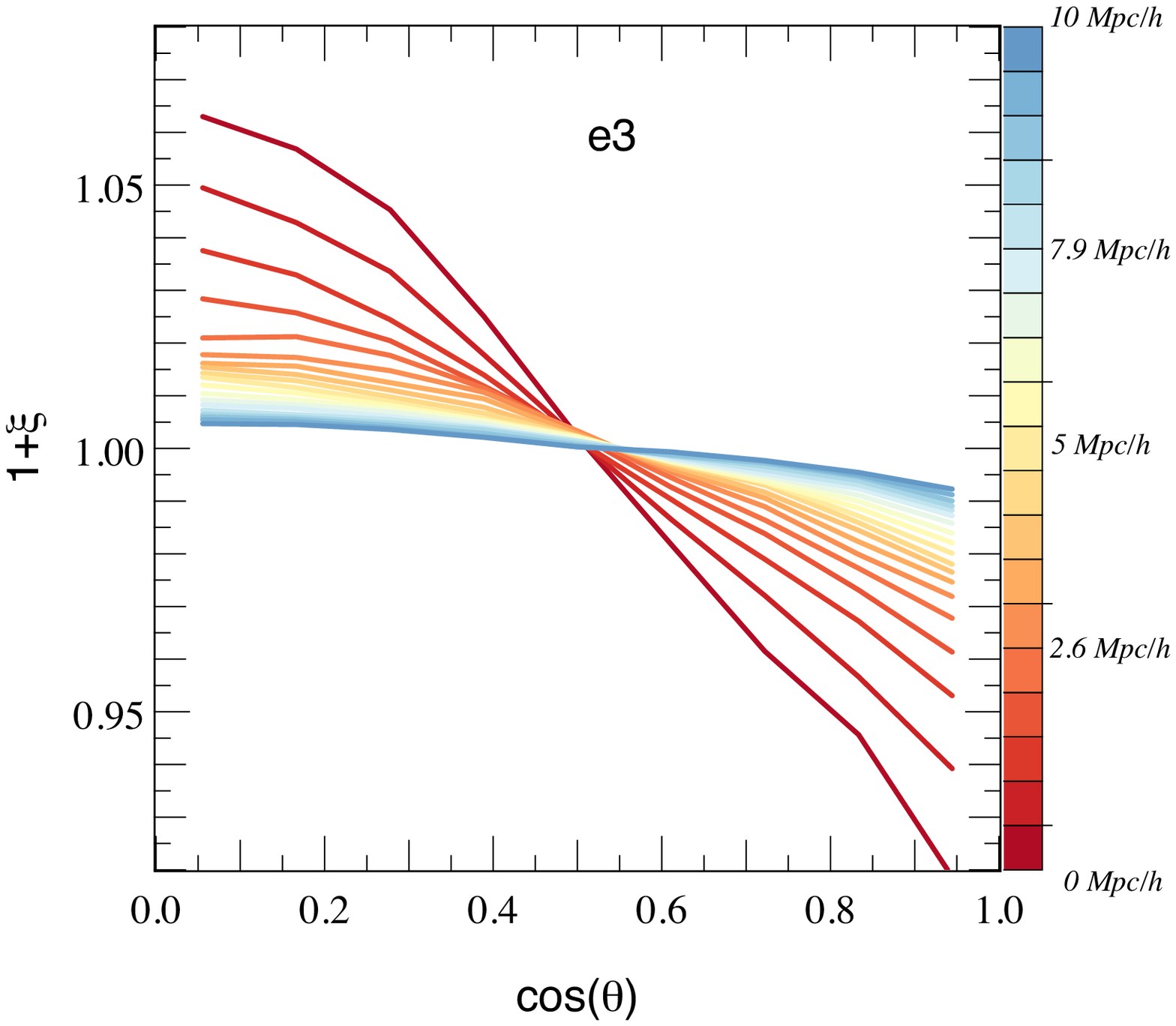}
\caption{
PDF of the cosine of the angle between the spin of galaxies and the tidal tensor (smoothed on 0.4 $h^{-1}\, \rm Mpc$) eigen-directions $\mathbf{e}_{1}$ (left panel), $\mathbf{e}_{2}$(middle panel) and $\mathbf{e}_{3}$ (right panel) at 2 locations separated by $r <10$ $h^{-1}\, \rm Mpc$ (comoving), the separation being colour-coded from red to blue. For the sake of readability, the one-sigma error on the mean estimating from 32 random resamples is displayed only on the left panel. The spins are more likely to be aligned with $\mathbf{e}_{1}$ and to a lesser extent with $\mathbf{e}_{2}$ at short distance in agreement with the one-point PDF shown in Fig.~\ref{fig:tidaltensor-gal}; this signal decreases when the separation increases as expected. This de-correlation is faster for $\mathbf{e}_{2}$ than for the other two eigen-directions, going from a few, to ten $h^{-1}\,\rm Mpc$.
\label{fig:2pt-spin-ei} 
}
\end{figure*}

\subsection{One-point cross-correlations}
\label{sec:onept-spin-tidal}

We begin with a measurement of the correlations between the spin and the eigen-directions of the tidal tensor at the same spatial position. In practice, we compute the cosine of the angle between the spin of the galaxies and the three eigen-directions of the local tidal tensor 
$\cos \theta=\mathbf{L} \cdot \mathbf{e}_{i}/\vert\mathbf{L}\vert$. The resulting histogram is shown
in Fig.~\ref{fig:tidaltensor-gal}.
The spin is preferentially aligned with the minor eigen-direction (i.e. the filaments) in agreement with the spin-filament correlations detected by \cite{dubois14} at redshift $z\sim1.83$. To a lower extent,  some alignment is found with the direction of the intermediate axis.

When galaxies are binned in mass (see Fig.~\ref{fig:tidaltensor-mass}), it appears that the most massive galaxies tend to have a spin lying in the plane ($\mathbf{e}_{2},\mathbf{e}_{3}$) perpendicular to the filaments, while the less massive galaxies have their spin aligned with $\mathbf{e}_{1}$. The transition occurs at stellar masses about $4\times 10^{10}\, \rm M_\odot$. We conclude that the spins of {\sl galaxies} are definitely influenced  by their surrounding environment
differentially with their mass.

Those findings follow very closely what can be found for DM halos, as detailed in Appendix~\ref{sec:halos} where a similar analysis for the alignment of the spin of the DM halos with the local tidal tensor is carried out. Fig.~\ref{fig:tidaltensor-DM} clearly exhibits the same  qualitative correlation as galaxies, namely  a transition at a halo mass $\sim 5\times 10^{11} \, \rm M_\odot$ from spins  aligned with $\mathbf{e}_{1}$, at low mass, to spins oriented in the plane $(\mathbf{e}_{2},\mathbf{e}_{3})$ at high mass. This is consistent with previous works based on pure dark matter simulations \citep[see in particular figure 3 in \citealp{codisetal12}]{calvoetal07,hahnetal07,pazetal08,zhangetal09,codisetal12,Libeskind13a,Calvo13}.

The   colour of galaxies is a quantity more  readily accessible  to observations. It is therefore of interest to see how different galaxy colours implies different IA as it would provide means on how to leverage this effect. Fig.~\ref{fig:tidaltensor-colors} displays the correlations between galactic spins, and the tidal field for different colours  as labeled (see section~\ref{sec:colours} for details about the extraction of galactic colours). The width of the colour bins has been chosen such that there is the same number of objects in each subset of galaxies. On average, the bluest galaxies (defined here by $u-r<0.78$) are more correlated with the tidal eigen-directions than the red galaxies ($u-r>1.1$ here). This can be easily understood from the fact that red galaxies are typically massive, while blue galaxies are often small-mass galaxies. At that redshift ($z\sim 1.2$), this implies that red galaxies correspond to objects around the transition mass, whereas blue galaxies are mostly aligned with $\mathbf{e}_{1}$. At lower redshift, we expect the population of massive galaxies perpendicular to $\mathbf{e}_{1}$ to increase, so that  red galaxies  become more correlated. Obviously, we should also keep in mind that applying additional selection cuts on the galaxy samples (mass, luminosity, etc) would change the level of correlation. For instance, red galaxies above $4\times10^{10}\, \rm M_{\odot}$ are more correlated than the whole population of red galaxies. 

\begin{figure}
\includegraphics[width=0.9\columnwidth]{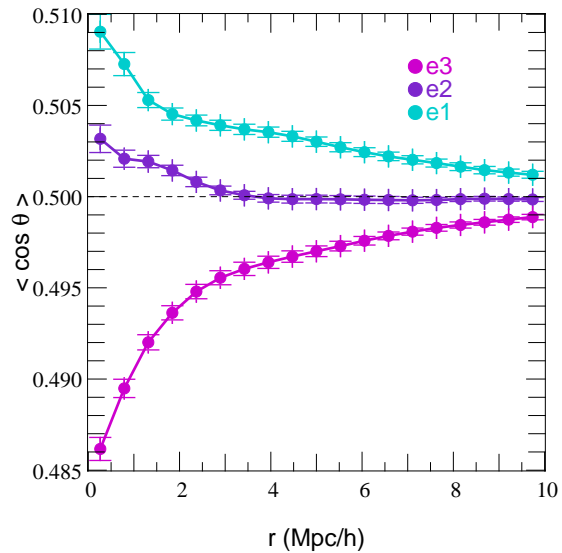}
\caption{Mean angle between the spin and the tidal eigen-directions as a function of the separation. Error bars represent the one-sigma error on the mean when cutting the statistics into 32 subsets. The spins are aligned with the minor direction (cyan) and to some extent with the intermediate eigen-direction (purple) at small distance and decorrelate on different scales. The alignment with $\mathbf{e}_{1}$ and $\mathbf{e}_{3}$ pervade on scales of about $10$ $h^{-1}\,\rm Mpc$ but is reduced to $\sim 3$ $h^{-1}\,\rm Mpc$ for the intermediate direction.
\label{fig:2pt-mean-spin-ei} 
}
\end{figure}

\subsection{Two-point cross-correlations}
\label{sec:twopt-spin-tidal}

While the aforementioned measurements have been performed at the same spatial location, it is also of interest in the context of weak lensing studies to quantify how this signal pervades when the separation  between galaxies increases.
Because the tidal field in the vicinity of a galaxy contributes also to the lensing signal carried by more distant galaxies, it is  clear that the spin -- tidal tensor cross-correlation is closely related to the so-called GI term in the weak lensing terminology. 
To address that question, we measure the correlations between the spins and the eigen-directions of the tidal tensor at comoving distance $r$. 
In practice, we compute for each pair of galaxies-grid cell (the tidal field being sampled on a $512^{3}$ cartesian grid) their relative separation and the angle between the spin of the galaxy and the three eigen-directions of the tidal tensor in the corresponding grid cell. We finally do an histogram of these quantities.
The results are shown in Fig.~\ref{fig:2pt-spin-ei}, which displays the PDF of the cosine of the angle between the spins and $\mathbf{e}_{1}$, $\mathbf{e}_{2}$, $\mathbf{e}_{3}$ as a function of the separation, and Fig.~\ref{fig:2pt-mean-spin-ei}, which shows on the same plot the mean angle with $\mathbf{e}_{1}$ (cyan), $\mathbf{e}_{2}$ (purple) and  $\mathbf{e}_{3}$ (magenta).
As expected, the spin and the tidal eigen-directions de-correlate with increasing separation. However, whereas the signal vanishes on scales $r > 3$ $h^{-1}\,Ê\rm Mpc$ for the spin to intermediate tidal eigen-direction correlation, it persists on distances as large as $\sim 10$ $h^{-1}\,Ê\rm Mpc$ for the minor and major eigen-directions of the tidal tensor.

\section{Spin-spin auto-correlations}
\label{sec:II}
In the previous section, we focused on the correlations between the spins and the tidal tensor eigen-directions as it is related to the ``GI term'' which is induced by correlations between the ellipticities and the cosmic shear.
We will now investigate the second source of IA that comes from the auto-correlations of the intrinsic ellipticities of galaxies. For that purpose, we   study first the spin-spin two-point correlation as a function of the galaxy pair separation (Section~\ref{sec:spin-spin}),
 before turning to the projected ellipticity two-point correlation function (Section~\ref{app:3to2}).

\subsection{3D spin-spin auto-correlations}
\label{sec:spin-spin}

\begin{figure*}
\includegraphics[width=0.66\columnwidth]{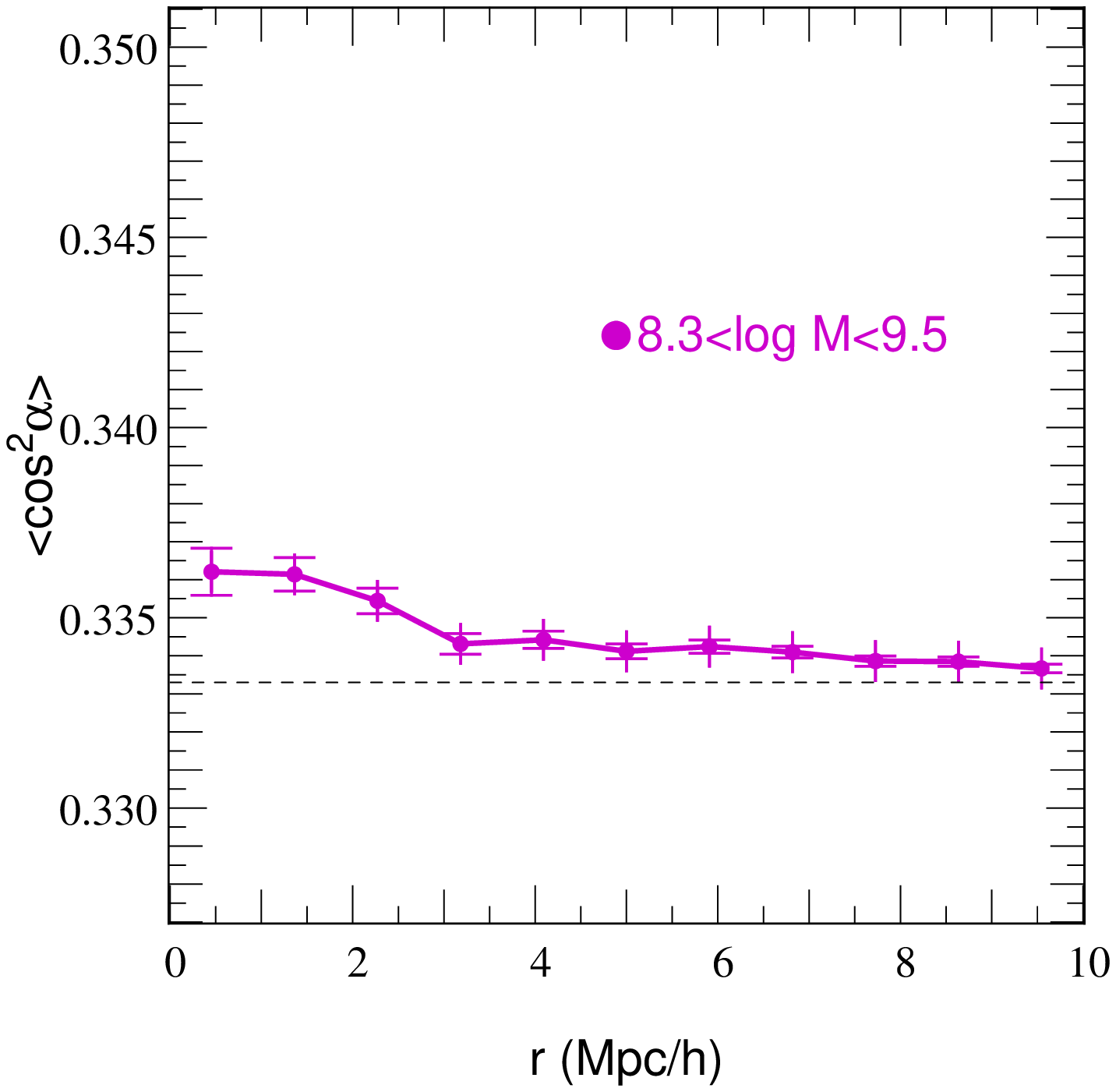}\includegraphics[width=0.66\columnwidth]{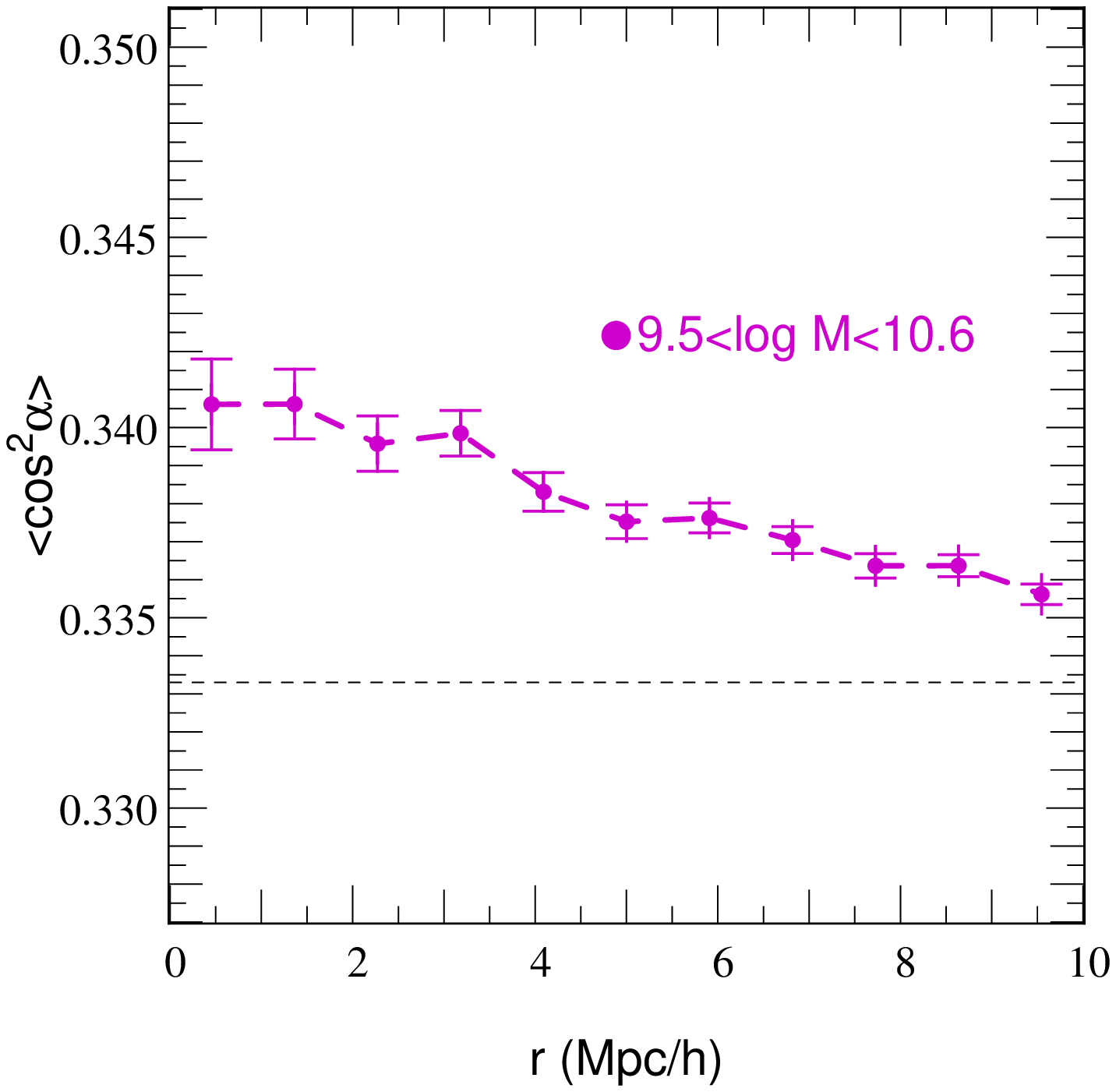}\includegraphics[width=0.66\columnwidth]{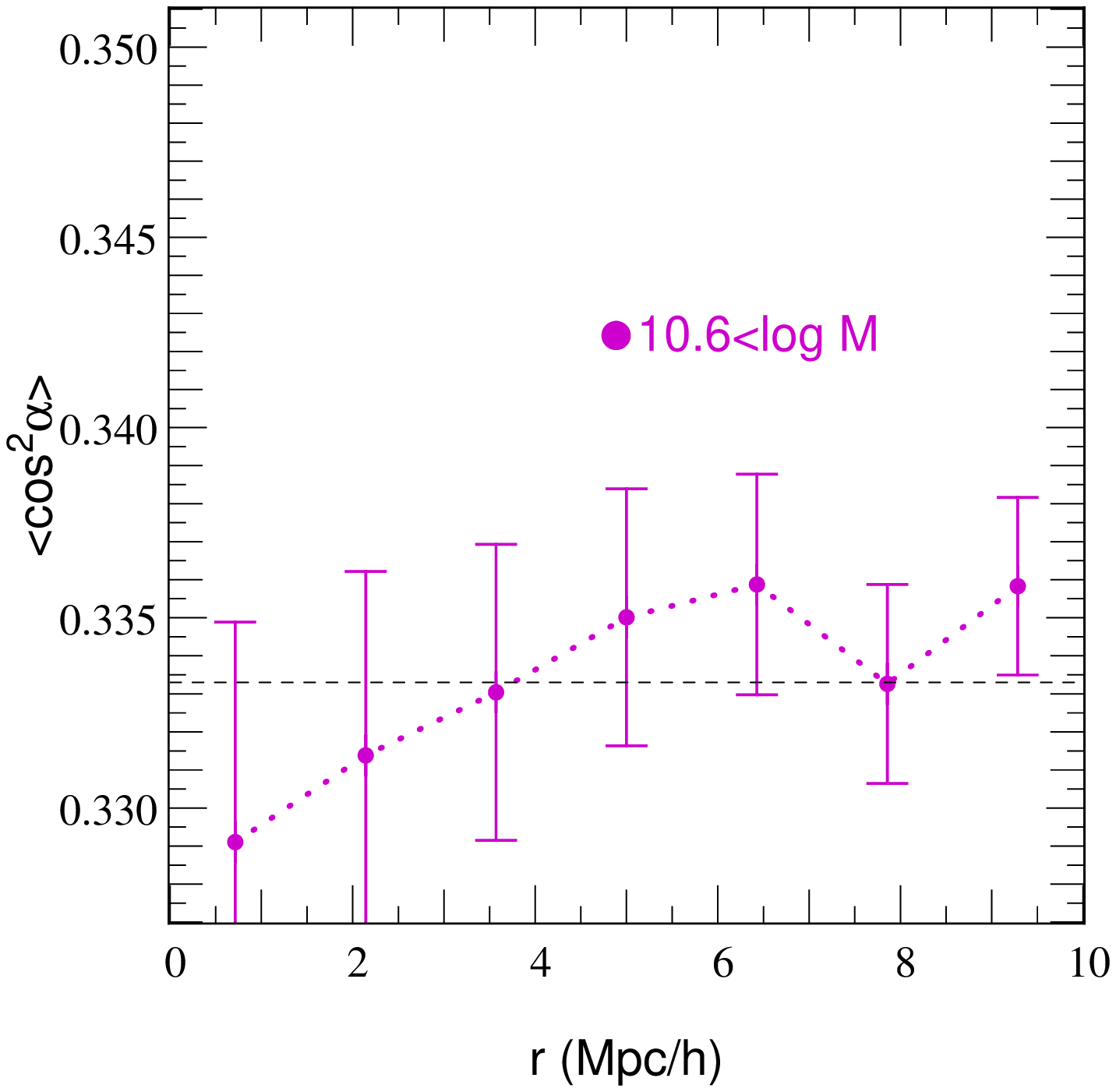}
 \includegraphics[width=0.66\columnwidth]{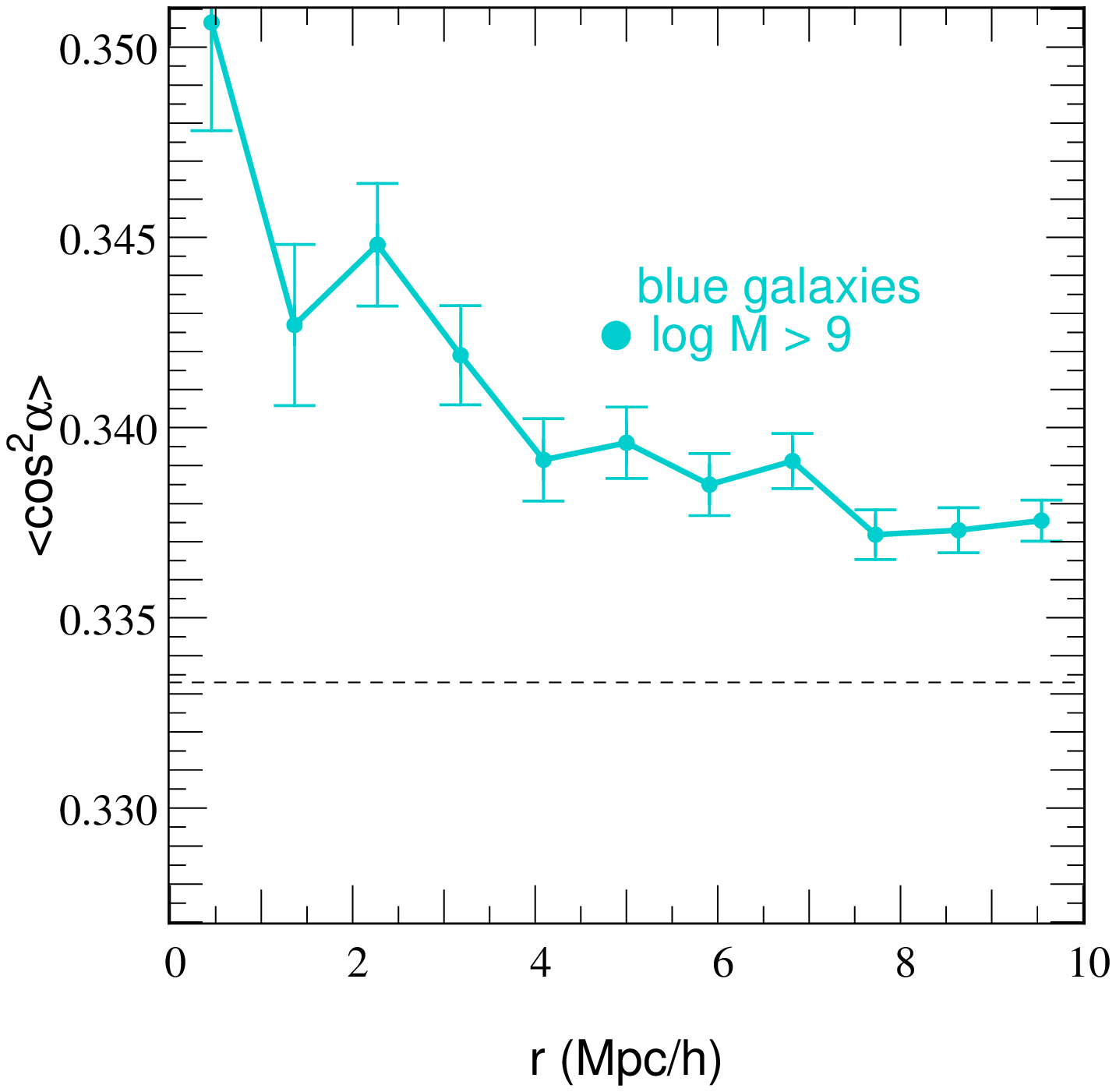}   \includegraphics[width=0.66\columnwidth]{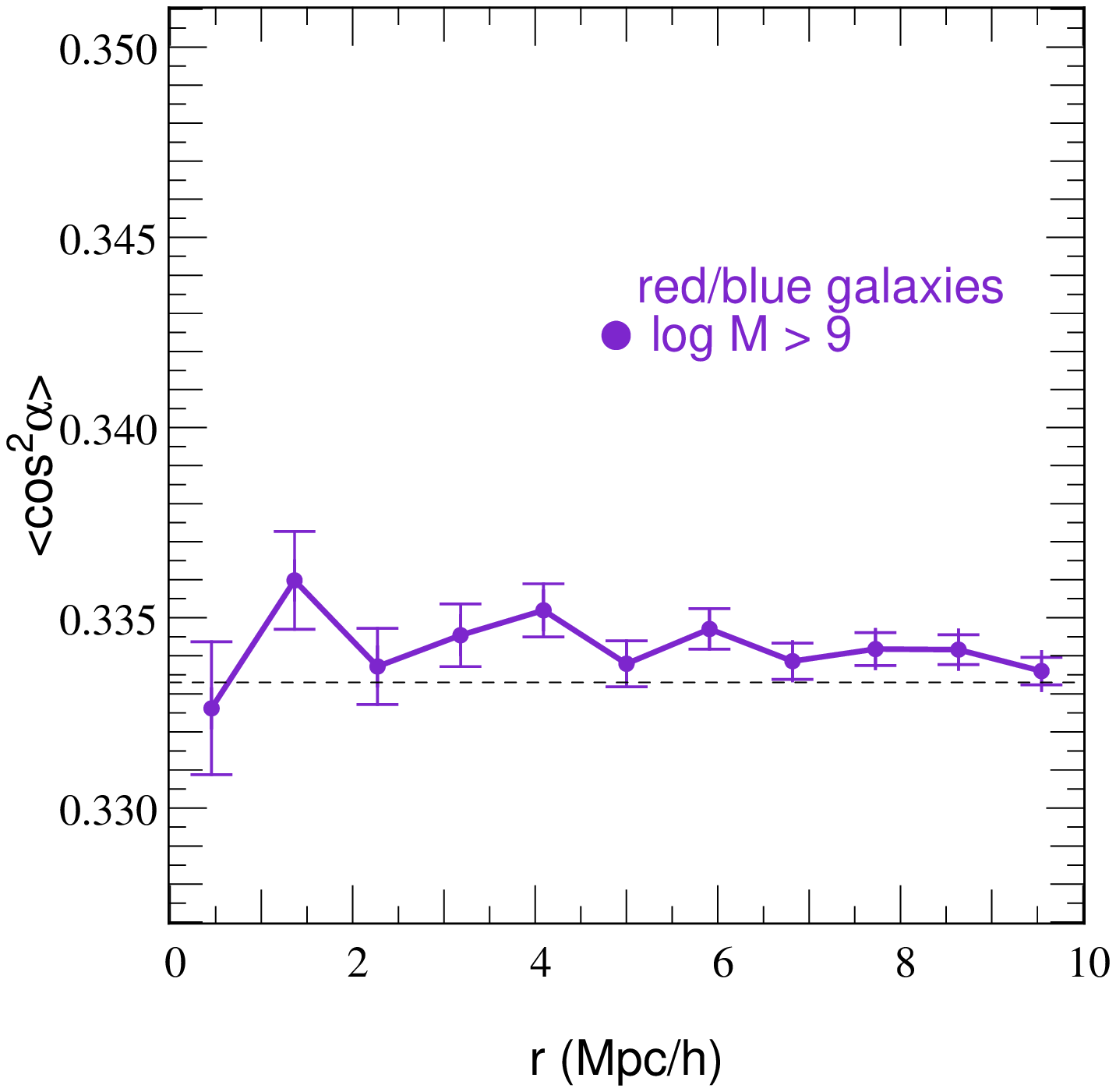}  \includegraphics[width=0.66\columnwidth]{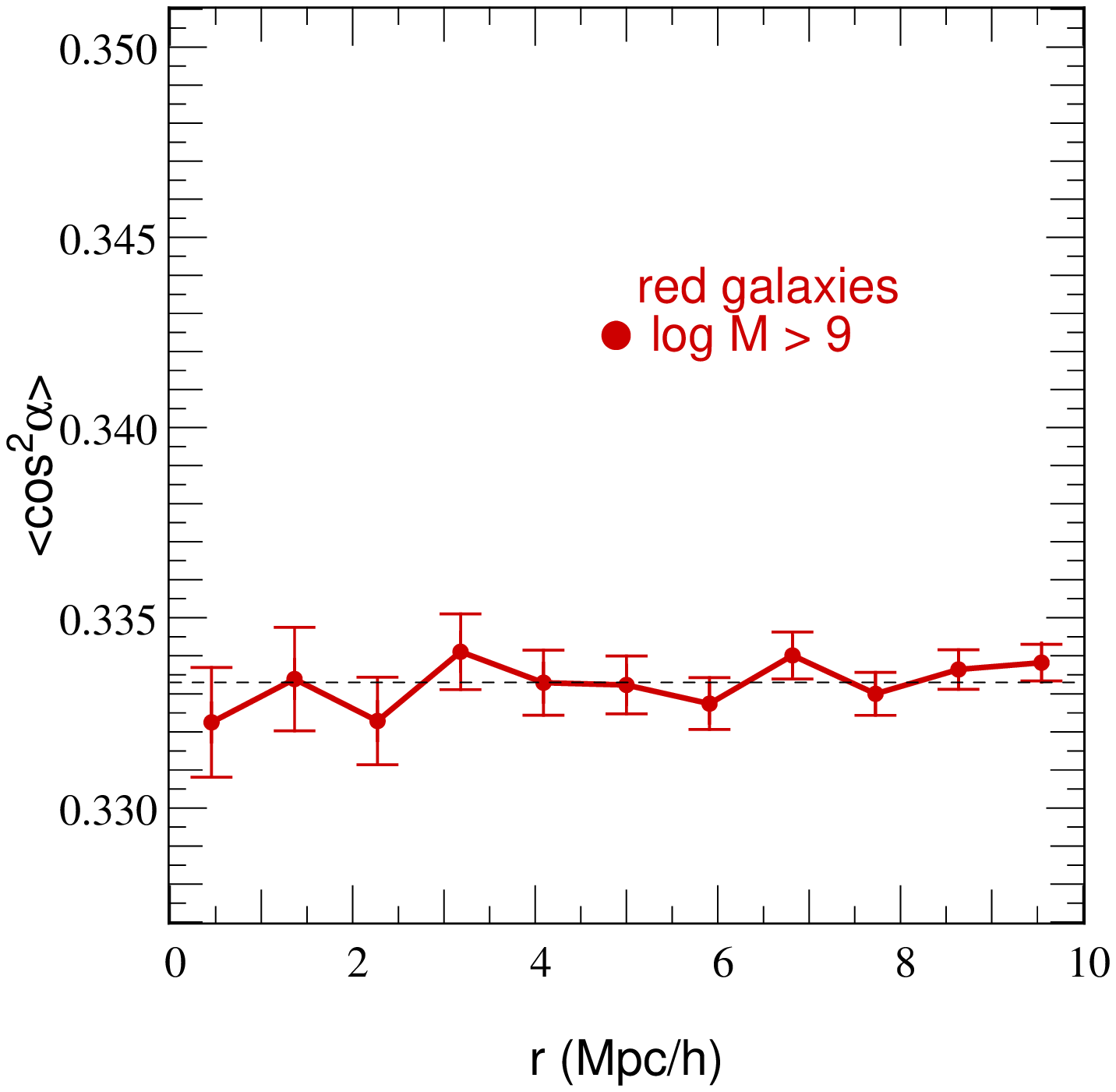}  \caption{3D spin-spin two-point correlation  function of galaxy  as a function of the comoving separation for a range of  stellar masses: $2\times10^{8}<M_{\rm s}<3\times10^{9}\, \rm M_\odot$ (top left panel), $3\times10^{9} <M_{\rm s}<4\times10^{10}\, \rm M_\odot$ (top center panel), and $M_{\rm s}> 4\times10^{10}\, \rm M_\odot$ (top right panel), or colours: blue galaxies i.e $u-r<0.78$ (bottom left panel), red galaxies i.e $u-r>1.1$ (bottom right panel), and the  cross correlations between blue and red galaxies (bottom center panel). Note that for blue and red galaxies we also apply a mass cut below $10^{9}M_{\odot}$ { (i.e 300 star particles)}. Error bars represent the error on the mean. Here we choose to display the mean square cosine between two spins (separated by the comoving distance $r$) as the polarity is not relevant to weak lensing studies. For a uniform random distribution, the expectation  is 1/3 (dashed line). Blue galaxies and small-to-middle mass galaxies are indeed correlated on the scale of the typical size of  filaments,  $\gtrsim 10 \, h^{-1}\, \rm Mpc$, whereas red and high-mass galaxies do not show a significant correlation.
 \label{fig:spin-spin} 
}
\end{figure*}

We begin with the auto-correlation of the direction of the spins as a function of the galaxy pair separation (in other words, the mean angle between the spin of two galaxies separated by a distance $r$). 
We select galaxies of different stellar masses: $2\times10^{8} <M_{\rm s}<3\times10^{9}\, \rm M_\odot$, $3\times10^{9}<M_{\rm s}<4\times10^{10}\, \rm M_\odot$ and $M_{\rm s}>4\times10^{10}\, \rm M_\odot$ and different colours.
For each pair of such galaxies separated by a comoving distance $r$, we measure the angle between their respective spin,
and compute the square of the cosine of this angle, $\cos^{2} \alpha$ (as the polarity is of no interest for weak lensing) in Fig.~\ref{fig:spin-spin}. 
Error bars represent the error on the mean  $\sqrt{\langle \cos^{4}\theta\rangle-\langle\cos^{2}\theta\rangle^{2}/ N}$.

We do not detect any significant spin correlation neither among red galaxies nor between red and blue galaxies. Conversely we measure a  significant spin correlation for blue galaxies out to  at least a comoving distance of 10 $h^{-1}\, \rm Mpc$. 
We also see that the correlation amplitude is strong for low and intermediate-mass galaxies.
The signal for the most massive galaxies or red galaxies is compatible with zero correlation at any distance.
The importance of grid-locking on these correlation   is estimated in { Section ~\ref{sec:grid}} where it is shown that it is not significant.

\subsection{2D ellipticity-ellipticity correlations}\label{app:3to2}
\begin{figure*}
\includegraphics[width=0.67\columnwidth]{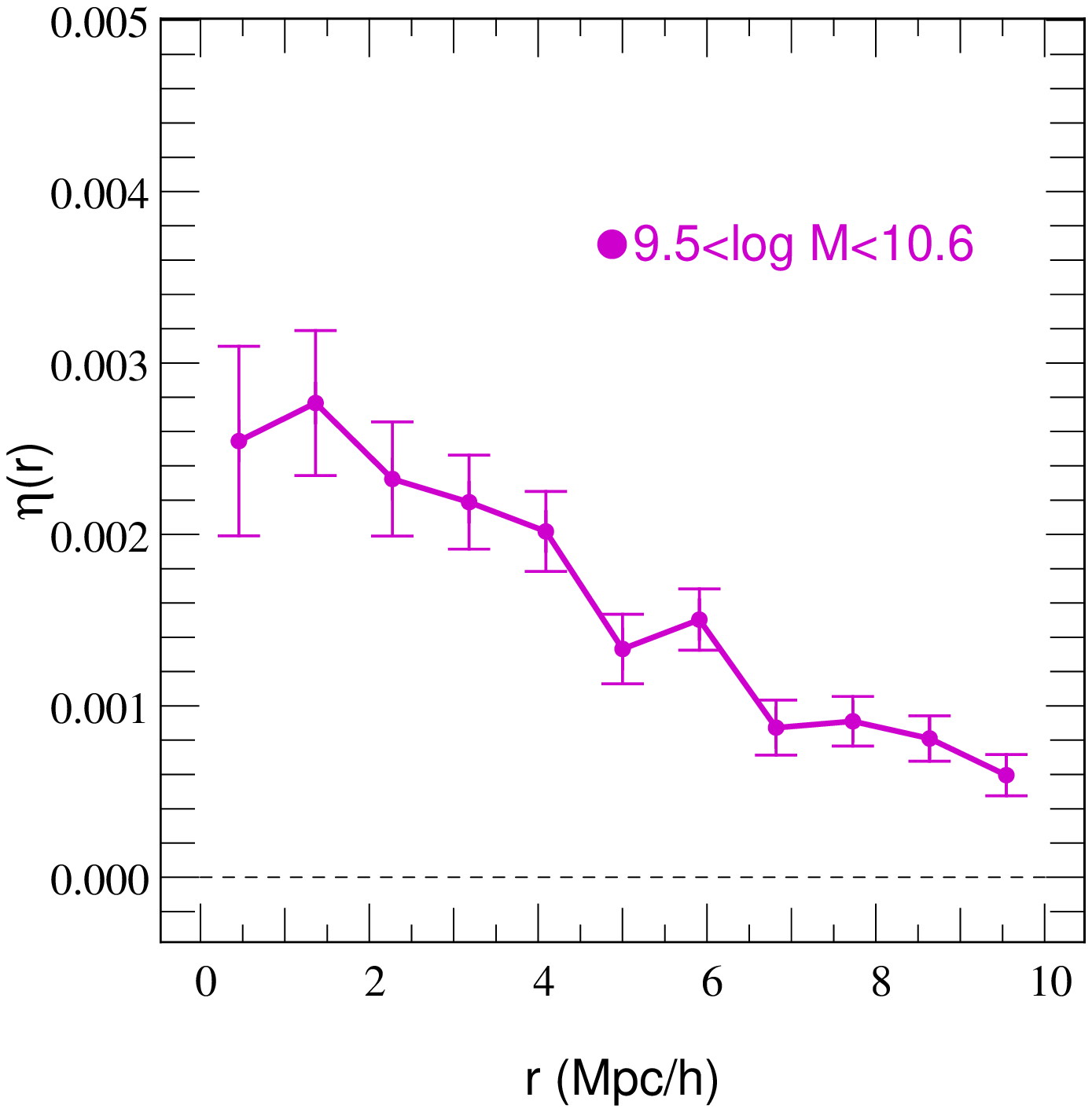}
\includegraphics[width=0.67\columnwidth]{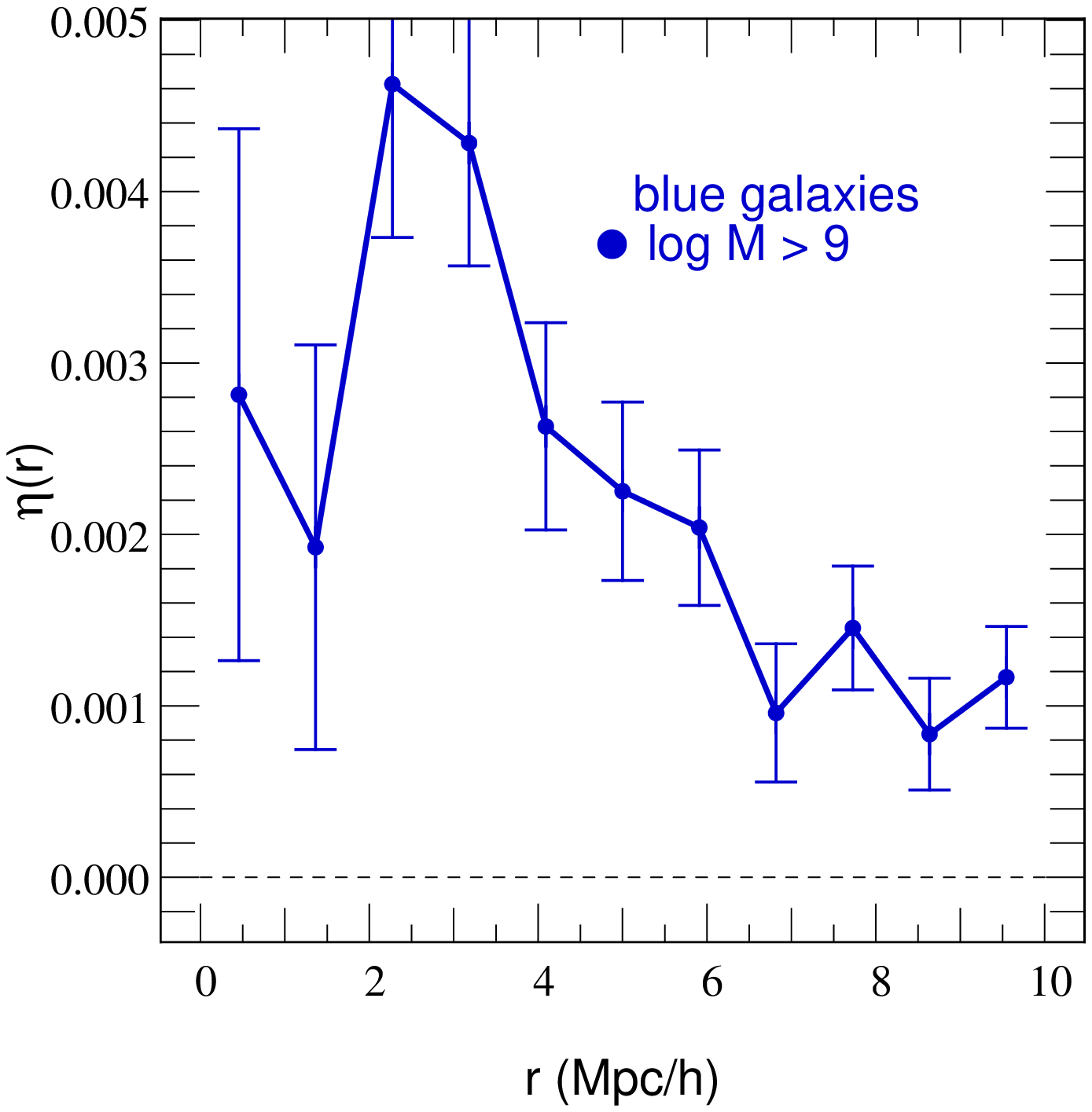}
\includegraphics[width=0.67\columnwidth]{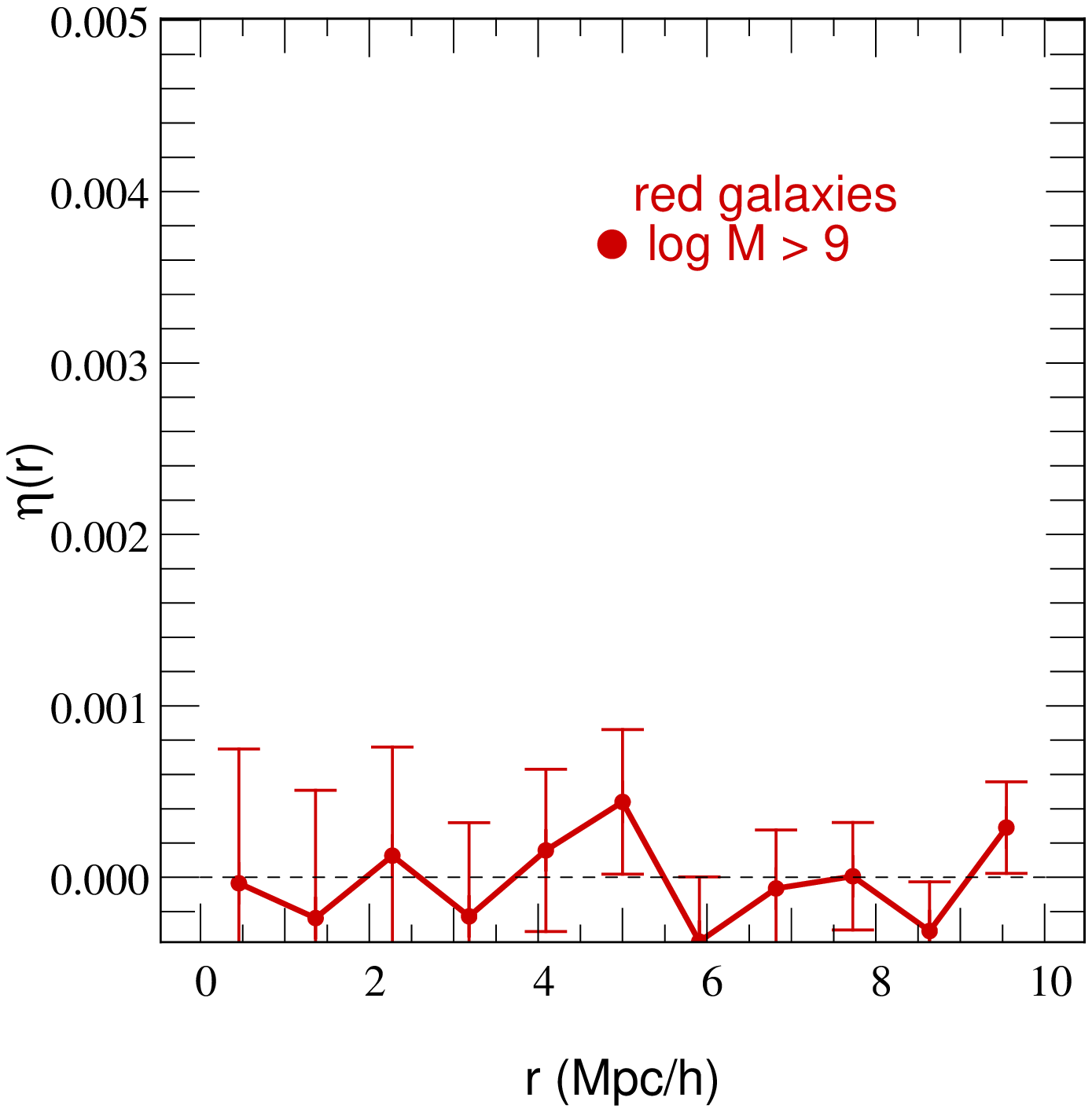}\\
\includegraphics[width=0.67\columnwidth]{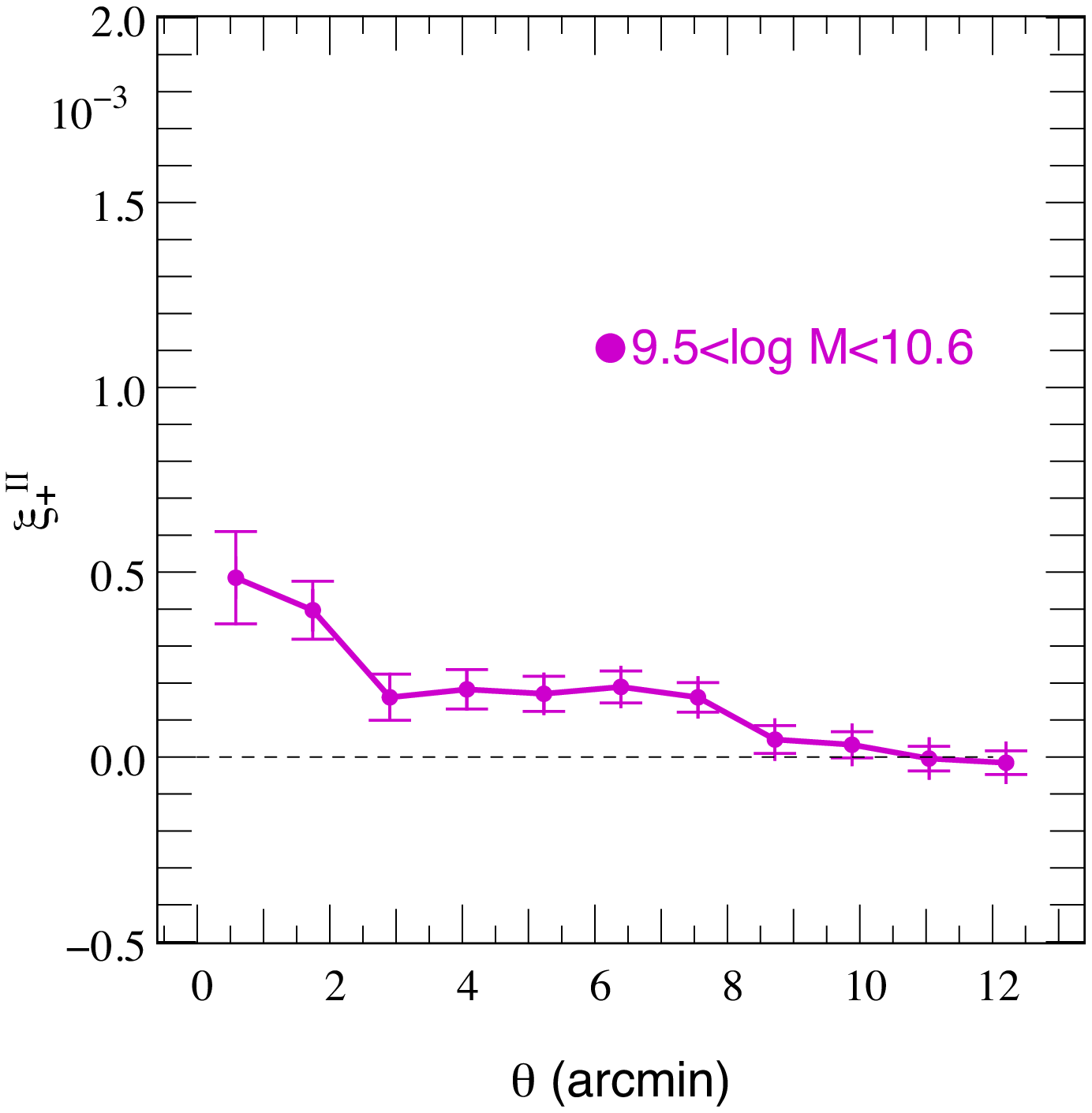}
\includegraphics[width=0.67\columnwidth]{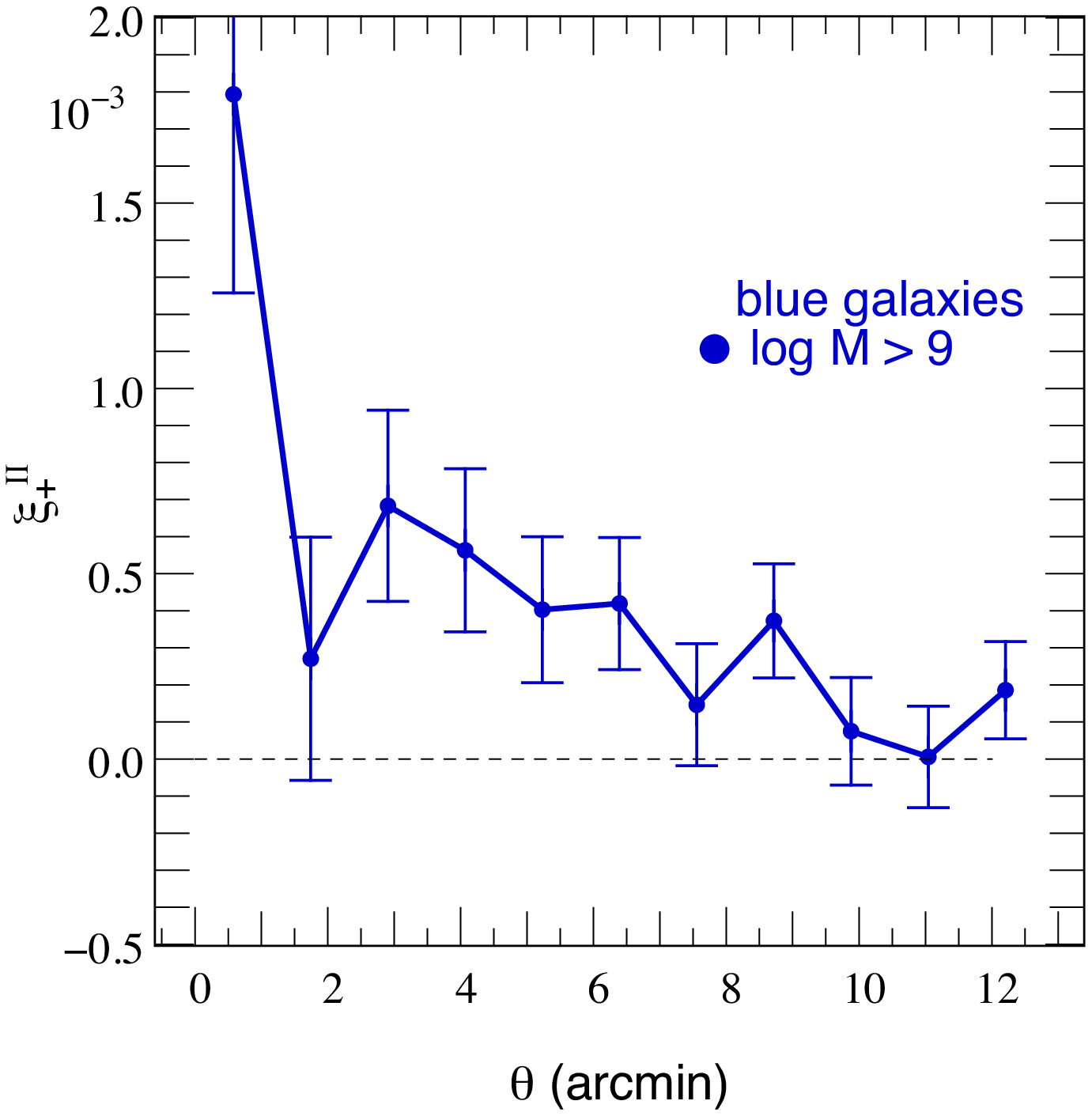}
\includegraphics[width=0.67\columnwidth]{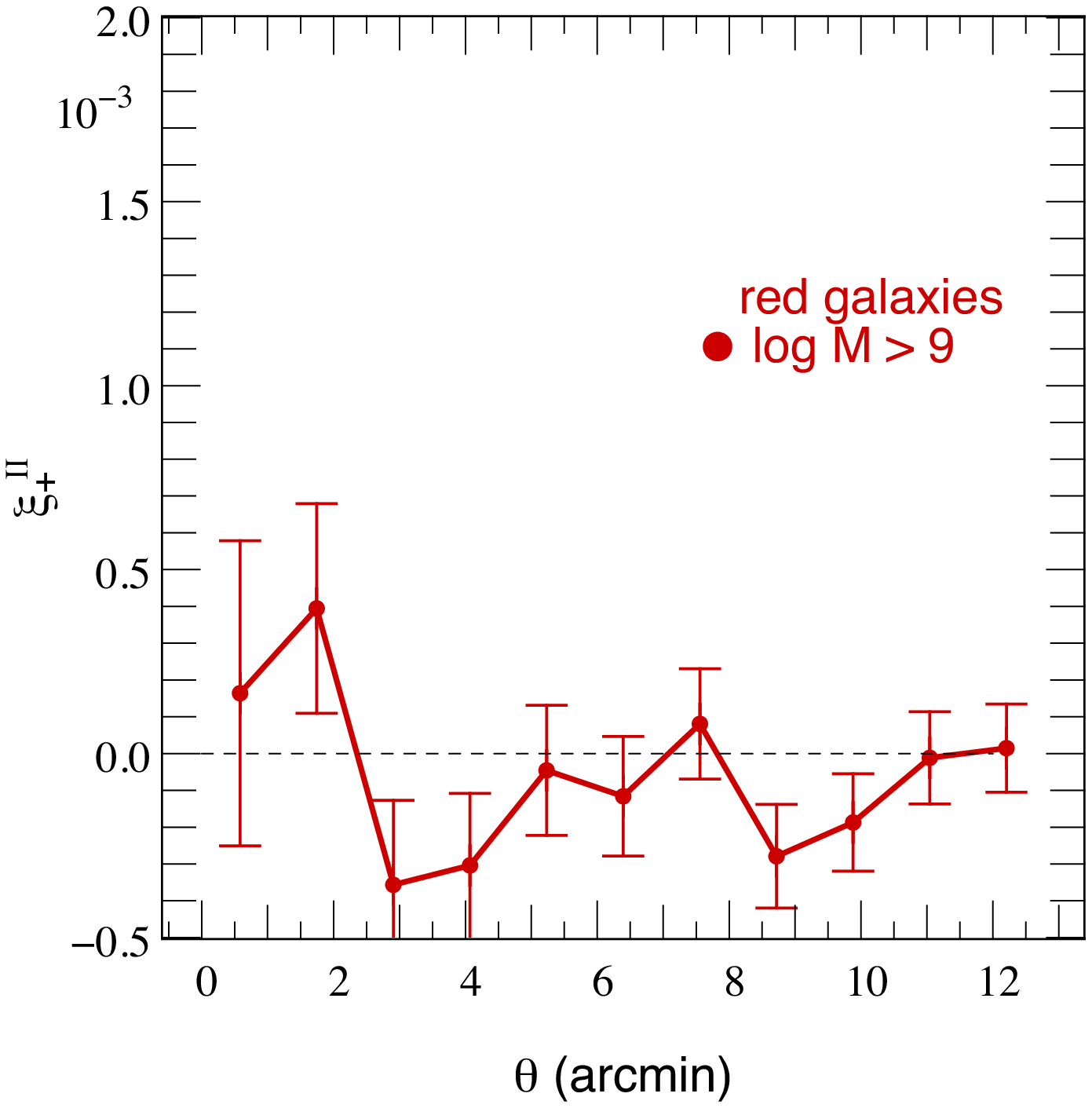}
\caption{Correlation functions of the projected ellipticity  $\eta(r)$ for the $\sim 58\,000$ middle-mass (left panels), $\sim 25\,000$ blue (middle panels) and $\sim 25\,000$ red (right panels) galaxies as a function of the comoving 3D separation $r$, in the top row, and as a function of the (projected) angular separation $\theta$ in the bottom row; this latter quantity being closer to observations, we call it $\xi_+^{\rm II}$.
Note the change in scale from one row to the other { and the additional mass cut below $10^{9}M_{\odot}$ (i.e 300 star particles)}.
\label{fig:eta}
}
\end{figure*}

The two quantities $\xi_{+}^{\rm II}$ and $\eta(r)$ defined in Section~\ref{sec:WL} are measured on our synthetic data and shown in Fig.~\ref{fig:eta}. The panels in the top row show our findings for $\eta(r)$ for three populations of galaxies where we choose a direction (here $(0.34,0.06,0.94)$ in cartesian coordinates) in the box different from the grid as the line-of-sight. Like the previous 3D analysis, there is a striking difference of behaviour between red and blue galaxies, the latter showing a strong correlation signal for the statistics of $\eta$ with a typical amplitude of $\sim 3\times 10^{-3}$ between 1 and 5 $h^{-1}\,\rm Mpc$. On the other hand, the correlation for red galaxies is compatible with zero.
We can therefore anticipate that blue galaxies at redshift $\sim 1.2$ should be affected by IA, leaving the possibility of a substantial contamination of the weak lensing signal. The bottom panels of Fig.~\ref{fig:eta} show the amplitude of the correlation function $\xi_{+}^{\rm II}(\theta)$ as a function of the angular galaxy pairs separation\footnote{Note that a $1\, h^{-1}\,Ê\rm Mpc$ comoving transverse distance corresponds  to an angular size of 1.3 arcmin at this redshift.}.
It is significant and comparable to the cosmic shear amplitude all the way to $\sim13$ arcmin.

\begin{figure*}
\includegraphics[width=0.67\columnwidth]{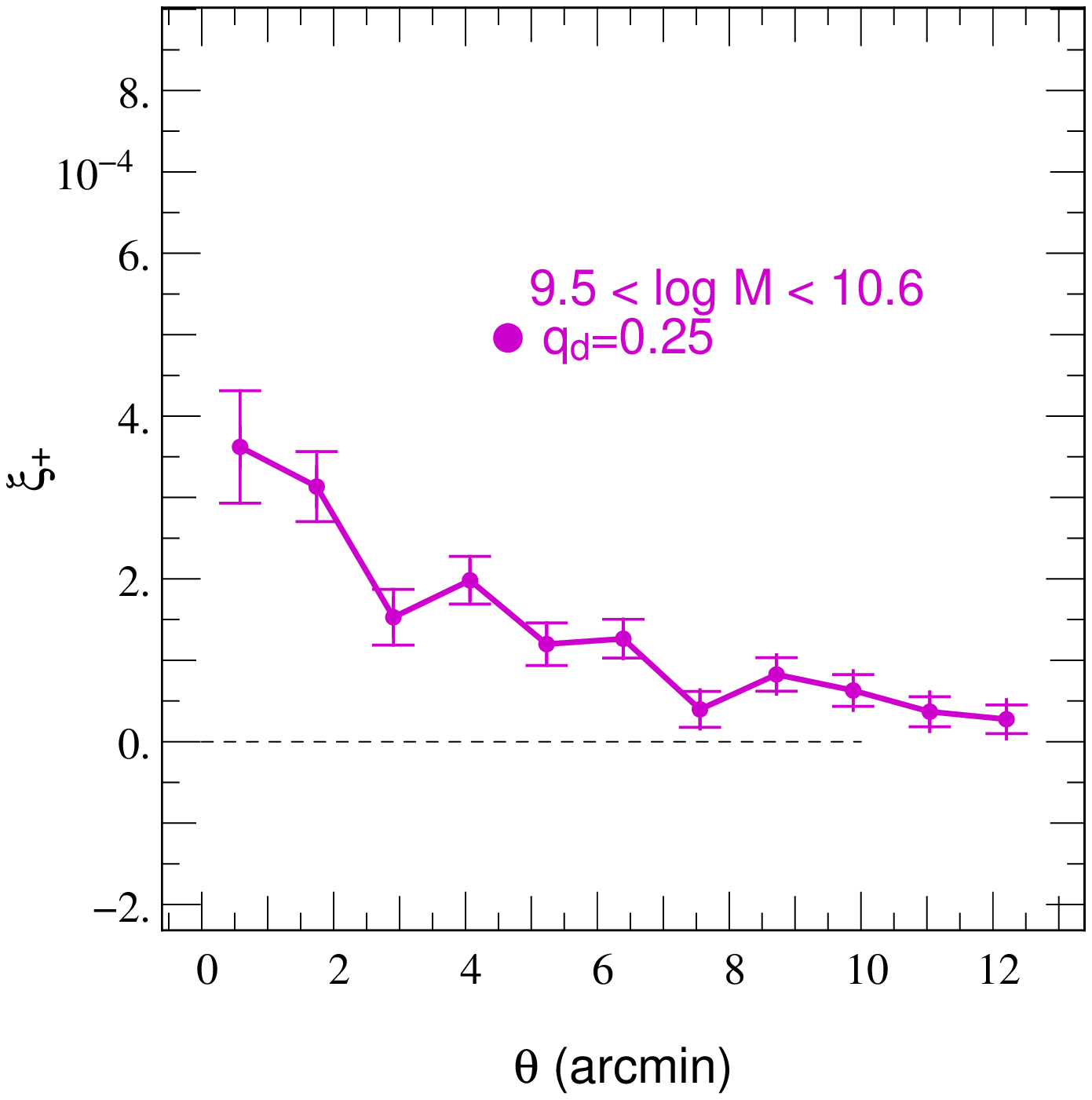}
\includegraphics[width=0.67\columnwidth]{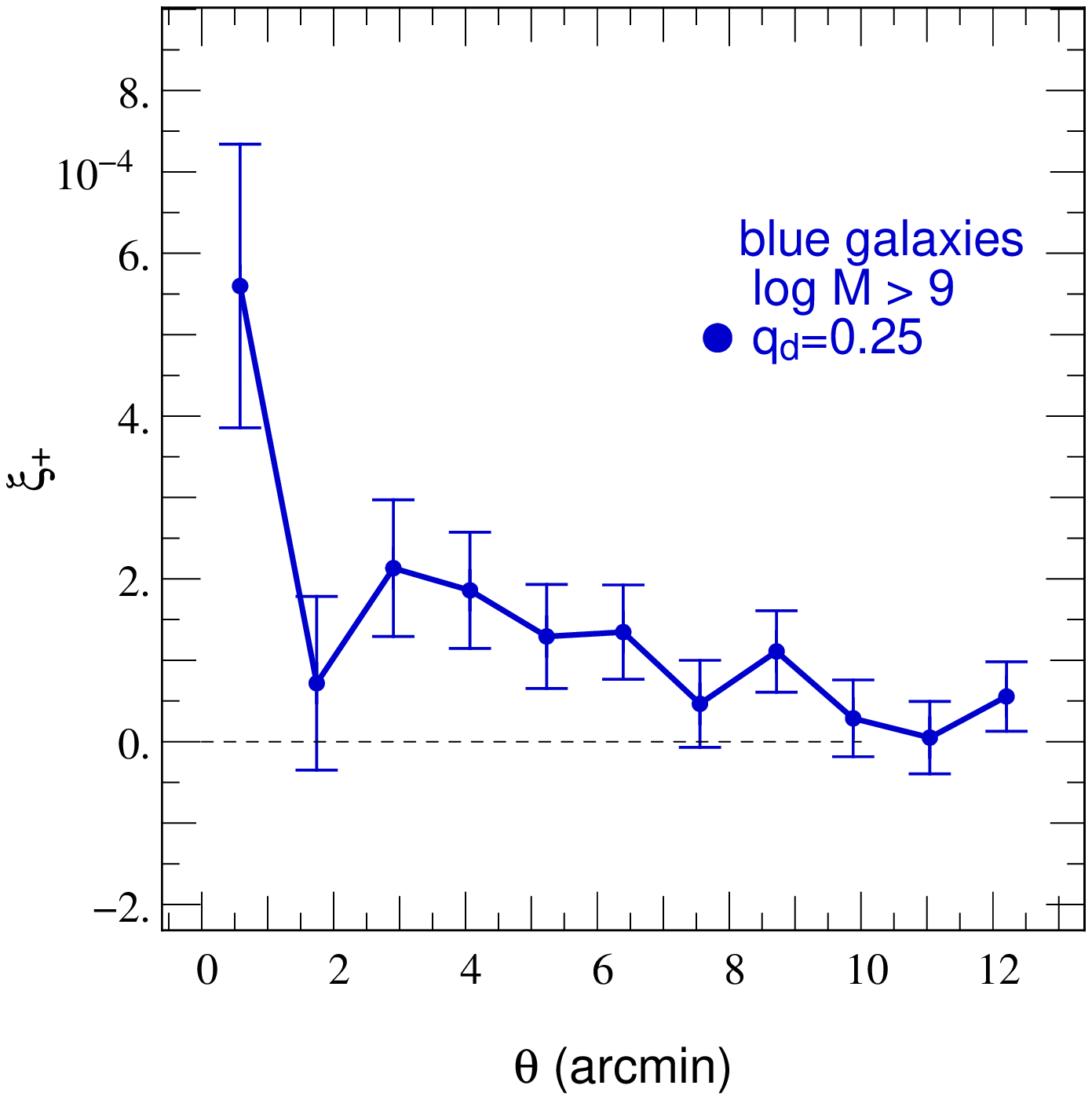}
\includegraphics[width=0.67\columnwidth]{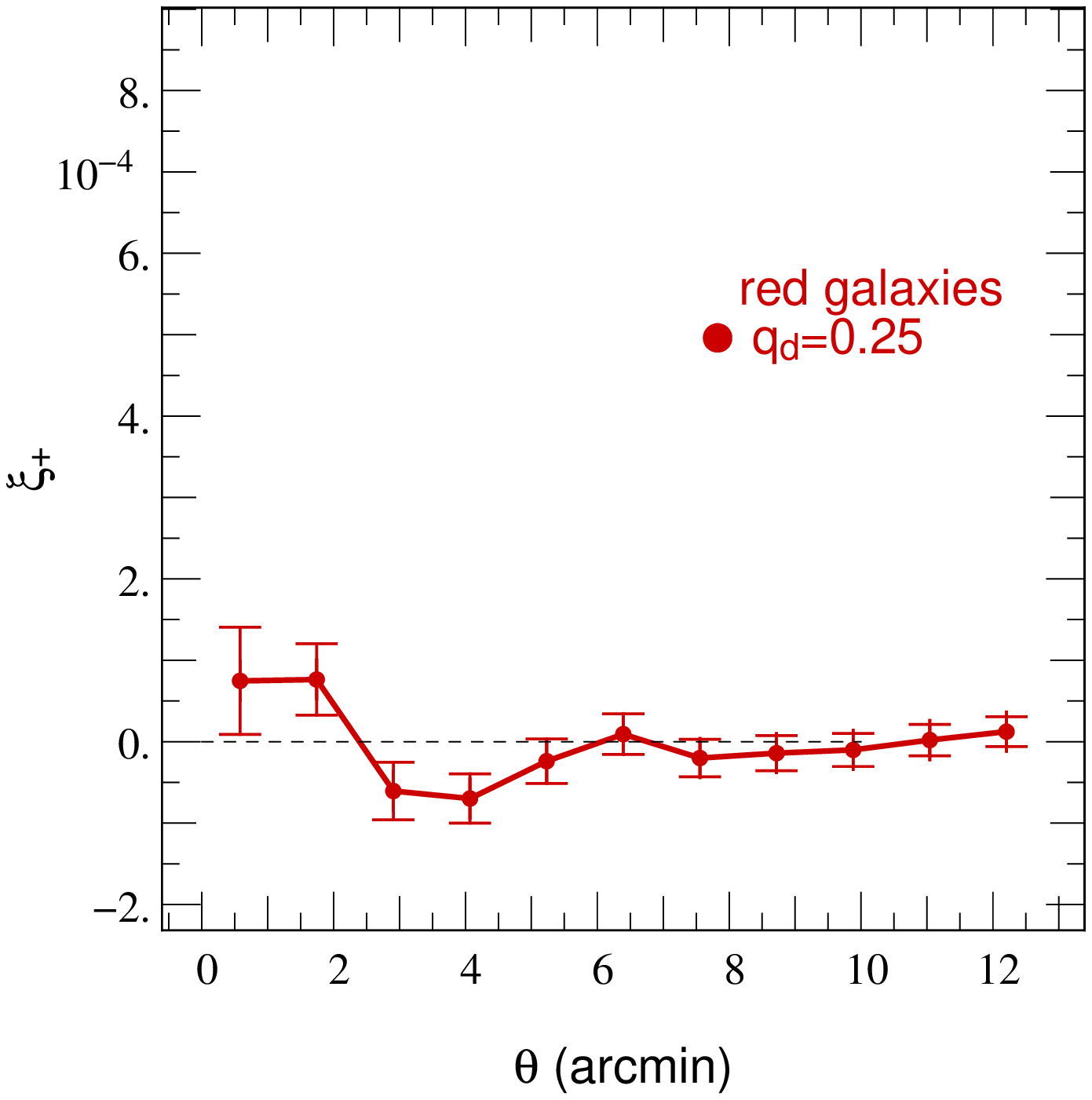}
\caption{{Same as bottom panels of Fig.~\ref{fig:eta} for thicker disks $q_{d}=0.25$.}
\label{fig:thicker}
}
\end{figure*}

Let us now compare our findings to the recent study of \citet{Joa++13b} (which is based on a semi-analytical model of galaxy formation). At first glance, we draw opposite conclusions, as we measure a strong level of IA for blue galaxies and no noticeable signal for red galaxies, whereas \citeauthor{Joa++13b} predict a strong alignment of red galaxies ($\eta\sim10^{-2}$ at $z=1.5$ and $r=1 \,h^{-1}\,\rm Mpc$) and a low level of alignment of blue galaxies ($\lesssim 3 \times 10^{-4} $ at $r=1  \,h^{-1}\,\rm Mpc$, compatible with zero correlation). Yet, several important differences should be emphasized:
\begin{itemize}
\item All our galaxies are assumed to be thin disks with a value $q_d=0$ in equation~(\ref{eq:L2q}) whereas \citeauthor{Joa++13b} explore two values $q_d=0.1$ and $q_d=0.25$ for late-type galaxies. Recall that our choice of $q_d=0$ will tend to maximise the amplitude of projected ellipticities. { We repeated the measurement of $\xi_+^{\rm II}$ assuming a value of $q_d=0.25$ which assumes quite a strong thickening of the disk. The net result shown in Fig.~\ref{fig:thicker}, is to decrease the normalisation of $\xi_+^{\rm II}$ for blue galaxies by a factor $\sim3$, to a typical amplitude of $\xi_+^{\rm II}\simeq 2\times10^{-4}$ between 1 and 5 $h^{-1}\,\rm Mpc$ at $z\sim1.2$ (but for intermediate-mass galaxies, the amplitude of the correlation remains qualitatively the same).}
\item Our distinction of red and blue galaxies may not correspond exactly to their early-type and late-type classification.  For instance, our red galaxies sample is dominated by low-mass red satellite instead of massive red central galaxies. But this effect cannot fully explain the difference.
\item To assign a spin to their late-type galaxies (somehow equivalent to our blue sample), \citeauthor{Joa++13b} used the results of \citet{Bett12} for the alignment between the DM halo spin orientation and the galaxy spin at redshift zero. These were obtained with several hydrodynamical zoom simulations in a $\sim 12.5\, h ^{-1}\, \rm Mpc$ or $\sim 20\, h ^{-1}\, \rm Mpc$ size volume. As shown in Fig.~\ref{fig:spin-gal-halo}, we find at redshift $z\sim1.2$ a somewhat weaker alignment between galaxy and halo spins than~\citet{Bett12}. Naively, this fact should produce weaker two-point galaxy spin-spin correlation than the \citeauthor{Joa++13b} results. However, \citet{dubois14} showed that the spin of galaxies are in fact as correlated with the large-scale filaments as their DM host halo. 
This is a consequence of cold flows 
that
advect efficiently the cosmic angular momentum all the way to galaxies at the center  of dark halos \citep{Pic++11,danovichetal11,kimmetal11,Stewart2013}. 
 {Therefore semi-analytic models, which chain the de-correlation
between the large-scale structure and the halo, and that between the halo and the disc galaxies, will most certainly under-estimate the correlation of these galaxies with the large-scale structure.}
\item  Conversely, the most likely explanation for the discrepancy of  the red population is twofold. 
First, the red galaxies here are objects around the transition mass. Hence, part of them have their spin aligned with the surrounding filament (below the transition mass), while the others have their spin perpendicular to it (above the transition mass). Therefore, on average no correlation is detected among that sample. Nevertheless, for larger-mass and lower redshift, this population is expected to become more strongly perpendicular to the filaments and thus more correlated. Again, one should bear in mind that any selection (mass, colour, luminosity etc.) may bias significantly the two-point ellipticity correlations.
Second, spins alone may not fully capture the shape of non-rotating, mostly triaxial early-type galaxies. We defer for future work a thorough analysis of ellipticity alignments with the inertia tensor of galaxies as a proxy for ellipticity. 

\end{itemize}

Note finally that the coherence of the large-scale structure is expected to be higher with increasing redshift. We may therefore speculate that, at later times, the intrinsic correlation of spins will decrease \citep[see also][]{Lee++08,Joa++13b}.
{
\section{Grid Locking effect on spin alignments}
\label{sec:grid}

\begin{figure}
\includegraphics[width=0.9\columnwidth]{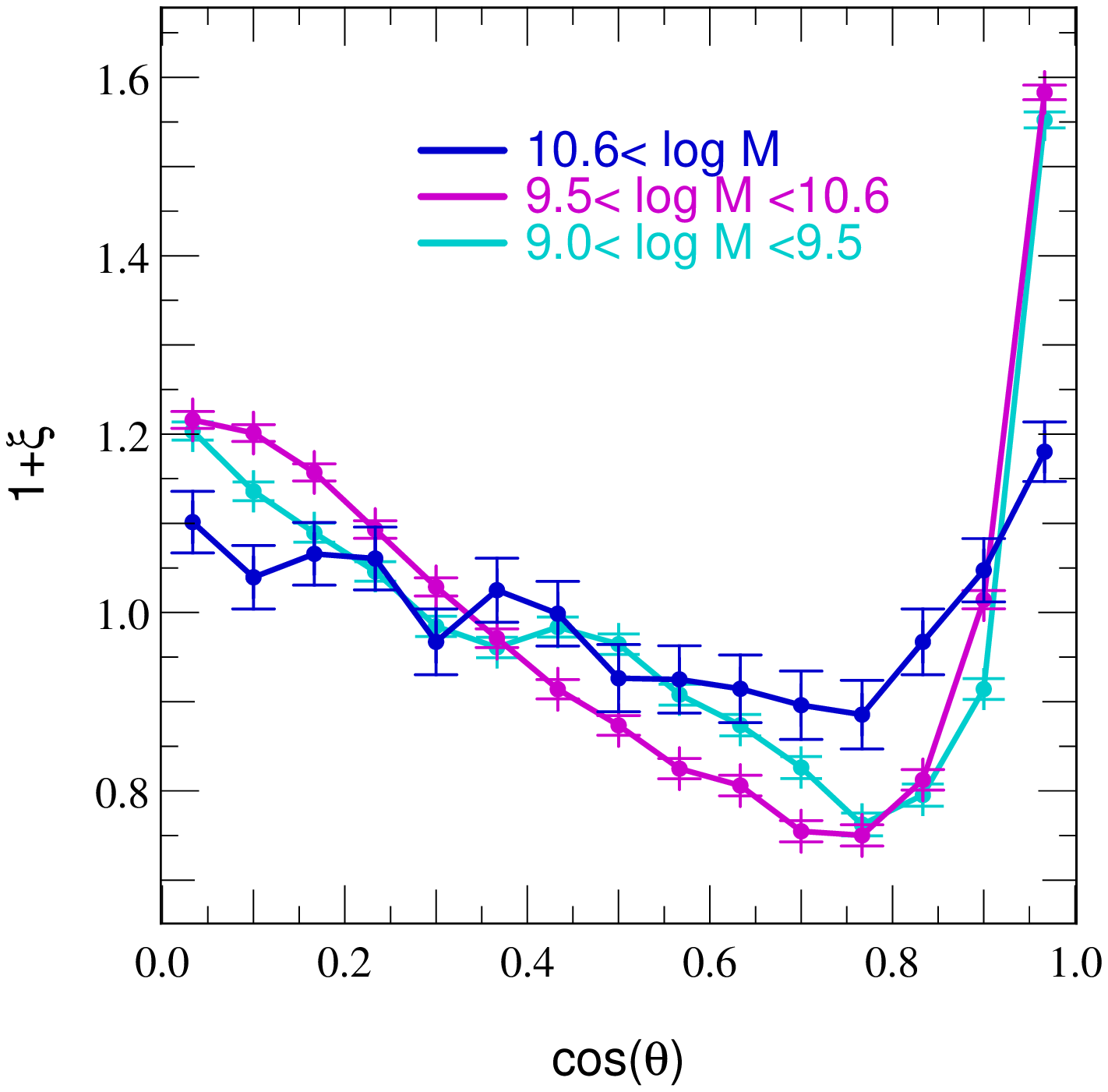}
\caption{{Excess probability of alignment, $\xi$, between the spins and the grid axes for different bins of mass. The most massive galaxies (blue solid line) are the least prone to grid locking. Note that the displayed plot is the mean on the PDF obtained for the x, y and z axes.}
\label{fig:GL} 
}
\end{figure}

\begin{figure}
\includegraphics[width=0.9\columnwidth]{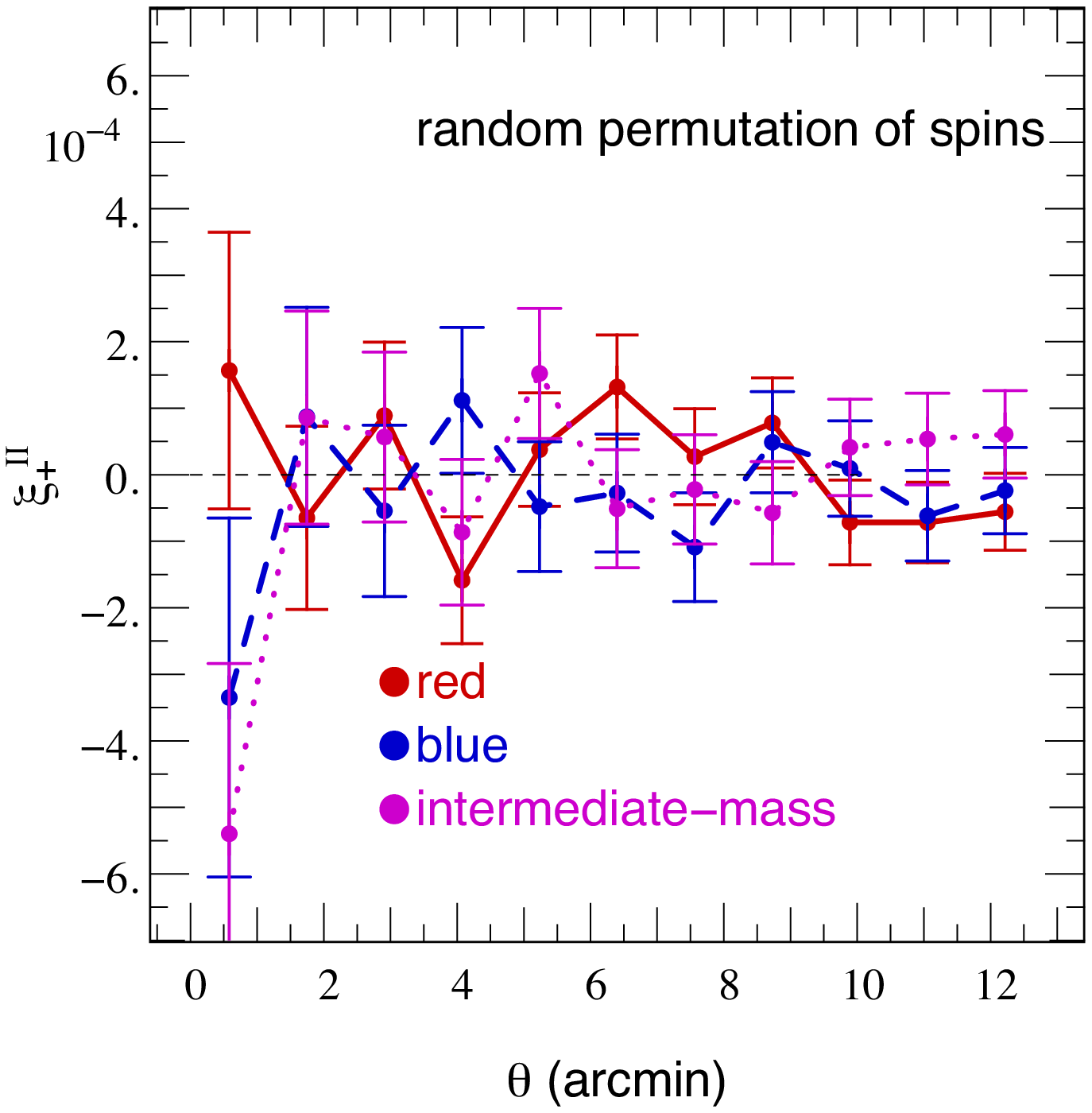}
\caption{$\xi_+^{\rm II}$, the two-point projected ellipticity correlation function for intermediate-mass, blue and red galaxies when spins are randomly swapped in the box. 
The three panels are statistically compatible with  non correlation, meaning that grid-locking does not introduce significant ellipticity correlations. This allows us to assume that the correlations detected in Fig.~\ref{fig:eta} are physical and not  due to grid-locking effects. 
\label{fig:xiplus-permutation} 
}
\end{figure}

Grid-locking effects are a concern for spin-spin correlation functions when using grid-based codes such as AMR (see also \cite{dubois14} for details about how spins are correlated with the cartesian grid). The alignments of the spins with the grid are shown in Fig~\ref{fig:GL}.
To investigate this effect on the two-point statistics studied in this work,
we compute the same statistics (spin-spin correlation functions and its 2D counterpart, $\xi_{+}^{\rm II}$) after a random permutation of the spins in the box.  This allows us to keep the same one-point distribution of spins on the sphere, including the corresponding level of grid-locking,
 but to remove the physical two-point correlations (the boxsize being much larger than the typical correlation length). Thus any  signal in the data -- relative to the  random permutation of spins, should be physical and not induced by grid-locking. 
Fig.~\ref{fig:xiplus-permutation} shows the result for different populations of galaxies (blue, red and intermediate mass). It appears that all those correlations are consistent with $\xi_{+}^{\rm II}=0$, implying that the results presented in the main text (e.g Fig.~\ref{fig:eta}) are not significantly biased  by grid-locking. The same  conclusion holds for spin-spin correlations and $\eta( r)$.

{
To be more precise,
let us assume that the spin of a galaxy in the simulation is the superposition of the ``real'' spin plus a numerical contribution coming from grid locking $\mathbf{s}=\mathbf{s_{s}}+\mathbf{s_{gl}}$ where we neglect normalisations (all spins are of norm 1 here). Then the two-point function of the measured spin of galaxies reads 
\begin{multline}
\left\langle \mathbf{s} (\mathbf{x})\mathbf{s}(\mathbf{x+r})\right \rangle=
\left\langle \mathbf{s_{s}} (\mathbf{x})\mathbf{s_{s}}(\mathbf{x+r})\right \rangle\\
+2
\left\langle \mathbf{s_{s}} (\mathbf{x})\mathbf{s_{gl}}(\mathbf{x+r})\right \rangle
+\left\langle \mathbf{s_{gl}} (\mathbf{x})\mathbf{s_{gl}}(\mathbf{x+r})\right \rangle
\end{multline}
Let us also assume that the spin contributions coming from grid locking do not depend on  spatial location. Then $\left\langle \mathbf{s_{s}} (\mathbf{x})\mathbf{s_{gl}}(\mathbf{x+r})\right \rangle$ and $\left\langle \mathbf{s_{gl}} (\mathbf{x})\mathbf{s_{gl}}(\mathbf{x+r})\right \rangle$ do not depend on $\mathbf{r}$ i.e those terms are constant. Therefore one can write
\begin{equation}
\left\langle \mathbf{s} (\mathbf{x})\mathbf{s}(\mathbf{x+r})\right \rangle=
\left\langle \mathbf{s_{s}} (\mathbf{x})\mathbf{s_{s}}(\mathbf{x+r})\right \rangle
+C\,.
\end{equation}
If one permutes all the spins in the box the physical two-point correlation of spins $\left\langle \mathbf{s_{s}} (\mathbf{x})\mathbf{s_{s}}(\mathbf{x+r})\right \rangle$ goes to zero if the box is large enough and $C$ remains $C$ as it does not depend on the spatial coordinates. We therefore measured this term in the simulation and found that $\vert C \vert \lesssim 10^{-4}$, which is below the statistical uncertainties currently plaguing our physical measurements. 
If grid locking does not depend on spatial location, we can conclude that the measured spin-spin correlation function is not significantly biased.

Another test that one can perform is to restrict the analysis to the galaxies that are less grid locked. We therefore measured the spin-spin correlation function for the most massive galaxies (that were shown to be the least sensitive to grid locking in Fig.~\ref{fig:GL}). The result is displayed in Fig.~\ref{fig:GL-masscut}. As the signal remains qualitatively the same (the amplitude is even larger), we conclude that grid locking can not be the main source of spin correlations measured in the simulation. Note that this test does not make any assumption on the physical origin of grid locking.

\begin{figure*}
\includegraphics[width=0.9\columnwidth]{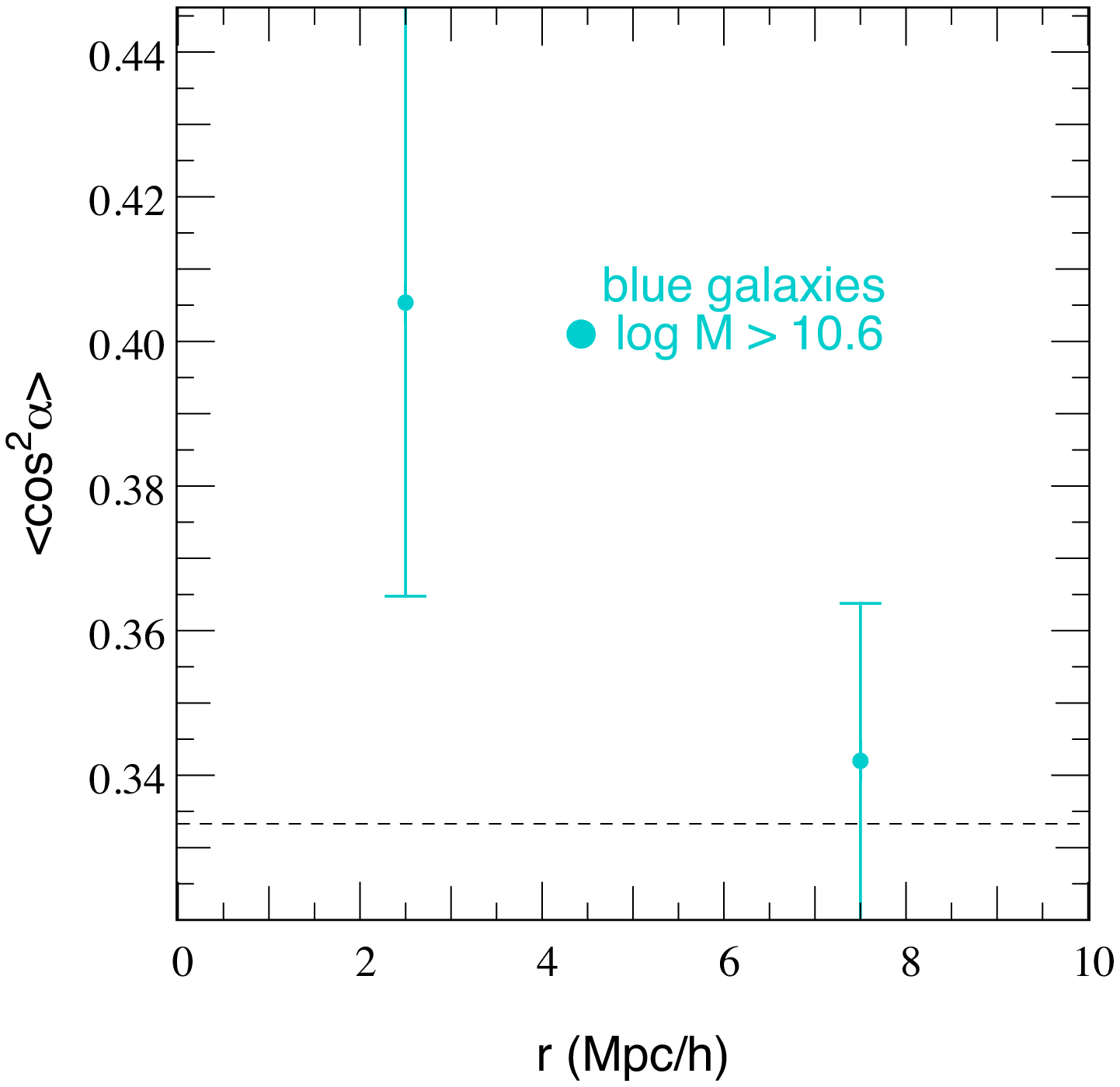}
\includegraphics[width=0.9\columnwidth]{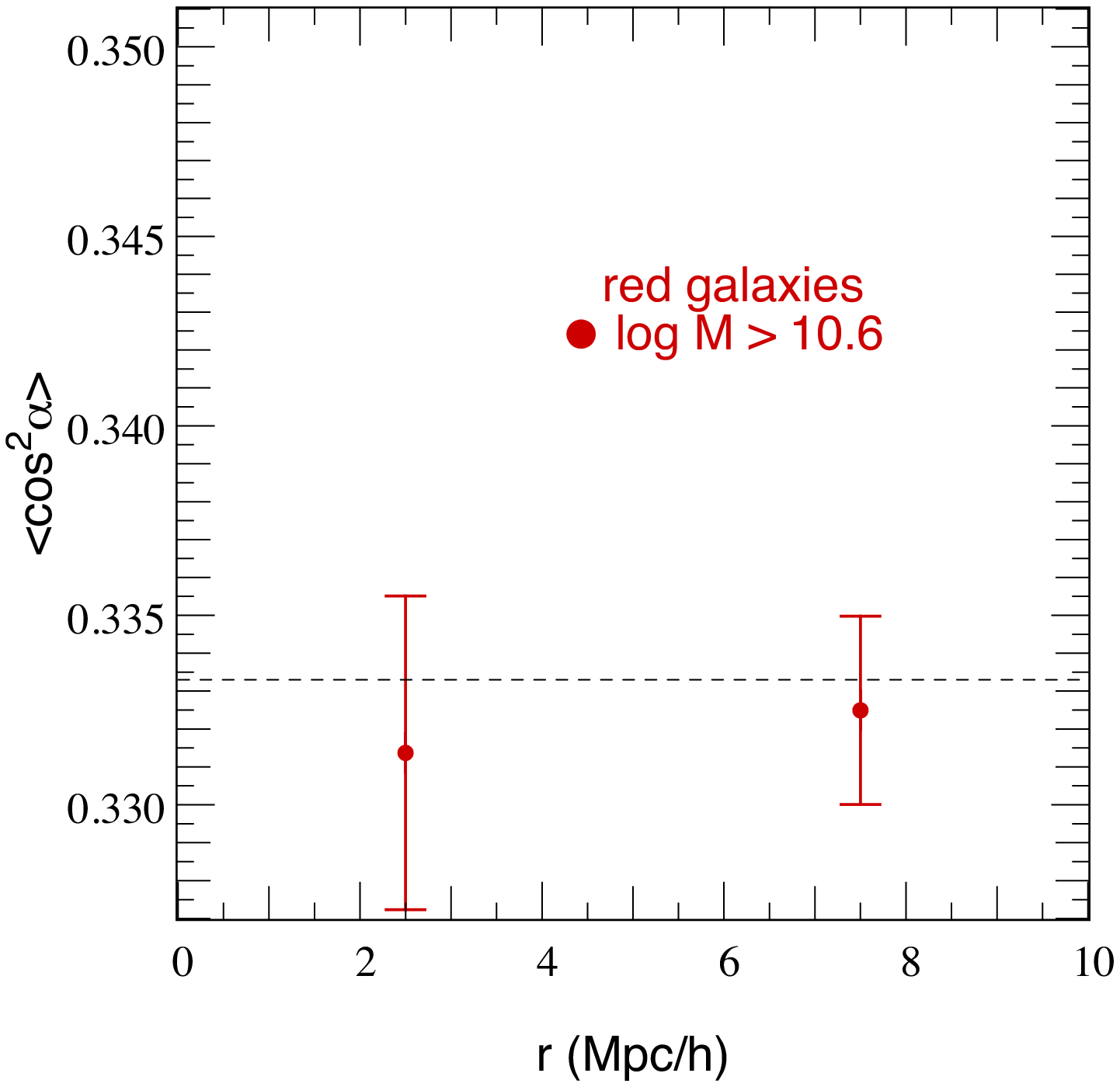}
\caption{3D spin-spin two-point correlation  function of galaxy  as a function of the comoving separation for blue galaxies i.e $u-r<0.78$ (left panel) and red galaxies i.e $u-r>1.1$ (right panel) with a mass cut below $4\cdot10^{10}M_{\odot}$. Same qualitative results as Fig.~\ref{fig:spin-spin}
are found. Hence, grid locking should not be the dominant source of the observed spin correlations.
\label{fig:GL-masscut} 
}
\end{figure*}

The only way to go beyond the tests proposed in this section would be to compare our results with similar cosmological hydrodynamics simulations performed using a technique which does not suffer from grid locking, but this 
is clearly beyond the scope of this paper.}
}

\section{Summary  and conclusion}
\label{sec:conclusion}

Using the \hagn simulation, we have shown
that low-mass galaxies tend to have their spin aligned 
with the local tidal field minor eigen-direction (the filamentary direction), whereas more massive galaxies ($M_{\rm s}> 4 \times 10^{10} \, \rm M_\odot$ at redshift one) have their spin mostly aligned with the major tidal eigen-direction (i.e perpendicular to walls and filaments).  
The corresponding two-point correlation decreases with the comoving separation out to scales as large as $\sim 10$ $h^{-1}\,\rm Mpc$, with a faster de-correlation for the spin-to-the-intermediate-axis direction ($\sim 3\, h^{-1}\,\rm Mpc$), a result consistent with Lagrangian theory in this context (Codis et al. in prep.).
Those results depend on the properties of galaxies, in particular on their mass and intrinsic colour. For instance, it was clearly found that at $z\simeq 1.2$ in the \hagn simulation, blue galaxies and intermediate-mass galaxies are significantly correlated with the gravitational tidal field whereas red and massive galaxies do not show any correlations.
We reach identical conclusions when studying the alignment of galaxy spins with one another, namely that
the spins of galaxies are also correlated on similar scales ($\sim 10$ $h^{-1}\,\rm Mpc$) and are similarly colour and mass-dependent.

We have also investigated how 
spin--spin correlations project into weak lensing observables like the shear correlation function $\xi_+$, 
these correlations being 
 cast into the so-called 
 II contributions to IA. As in 3D, a $\xi_+^{\rm II}$ correlation at a level of a few $10^{-4}$ is found for blue and intermediate-mass galaxies out to separations of $\gtrsim 10\arcmin$ for sources at redshift $\sim 1.2$.
The results for blue galaxies are in broad agreement with the recent work of \citet{Joa++13b}, who combines observational results on IA from the COSMOS survey and predictions from semi-analytical models applied to DM-only simulations. 
{ The effect of grid locking on the two-point functions were shown to be subdominant. However, to go beyond this qualitative statement and accurately quantify the effect of grid-locking, one would need 
to perform a similar simulation using a numerical technique insensitive to this specific systematic error, such as SPH
or unstructured mesh.}

Presently, the ``spin-gives-ellipticity'' prescription allows one to quantify the new insights that large volume hydrodynamical cosmological simulations bring to the issue of IA. For instance, the large-scale coherence of gas motions advected all the way to the center of galaxies through cold flows regardless of the DM behaviour can uniquely be captured by such  simulations~\citep{kimmetal11}. Large-scale dynamics imprint their coherence and  morphology (filaments, walls, voids) onto the spin of galaxies. This complex topology is likely to have an even more prominent impact on higher-order statistics beyond the shear two-point correlation function. Attempts to capture such effects with simple halo occupancy distribution prescriptions may therefore fail at high redshift ($z \gtrsim 0.8$), which is the place where galaxies carry more cosmological lensing signal and  is also, to large extent, the population of sources targeted by future surveys like Euclid or LSST. The challenge for simulations is to cover large cosmological volumes while preserving a sufficient resolution so that baryonic physics (star formation, feedback processes, etc) is correctly treated.

When the \hagn simulation  reaches redshift zero, we will be in a good position to compare our findings with existing observations. In order to get a good match for massive red galaxies, we will certainly adopt a different ansatz for our recipe -- currently based on a thin disk approximation -- and  use directly the resolved shape of massive galaxies as a proxy for the projected ellipticities. However, this concerns only a small fraction of the galaxies that made up the typical weak lensing catalogues of background sources. 
Once the  \hagn light-cone is completed, we will  estimate more realistic galactic shapes, taking full account of the spectral energy distribution of young and old stars (giving a non trivial weight to the relative contribution of the disk and the bulge) into a well chosen rest-frame filter (e.g. using the broad Euclid VIS band), and more precisely mimicking observational selection effects (i.e. a flux-limited sample of background sources will capture specific populations of sources at a given redshift). This will allow us to quantify the amount of contamination from IA expected in real surveys, and possibly mitigate their nuisance by selecting galaxies that are less prone to IA based on colour.

\section*{Acknowledgments}
This work has made use  of the HPC resources of CINES (Jade supercomputer) under the allocation 2013047012 and 2014047012 made by GENCI.
Part of the analysis was performed on the DiRAC facility jointly funded by STFC and the Large Facilities Capital Fund of BIS.
This work is partially supported by the Spin(e) grants {ANR-13-BS05-0005} of the French {\sl Agence Nationale de la Recherche}
and by the ILP LABEX (under reference ANR-10-LABX-63 and ANR-11-IDEX-0004-02).
The research of YD
has  been supported at IAP by  ERC project 267117 (DARK) hosted by Universit\'e Pierre et Marie Curie - Paris 6.
We thank  S. Rouberol for running  smoothly the {\tt Horizon} cluster for us,  and
D.~Munro for freely distributing his {\sc \small Yorick} programming language and opengl interface (available at {\tt http://yorick.sourceforge.net/}).
SC and CP thanks Lena for her hospitality during the course of this work.
VD would like to thank the Institut d'Astrophysique de Paris for hospitality during the completion of this work,
and acknowledges support by the Swiss National Science Foundation.
JD and ASÕs research is supported by funding from Adrian Beecroft, the Oxford Martin School and the STFC.

\bibliographystyle{mn2e}
\bibliography{author}

\appendix
\section{Dark matter halos}
\label{sec:halos}

\begin{figure*}
\includegraphics[width=0.68\columnwidth]{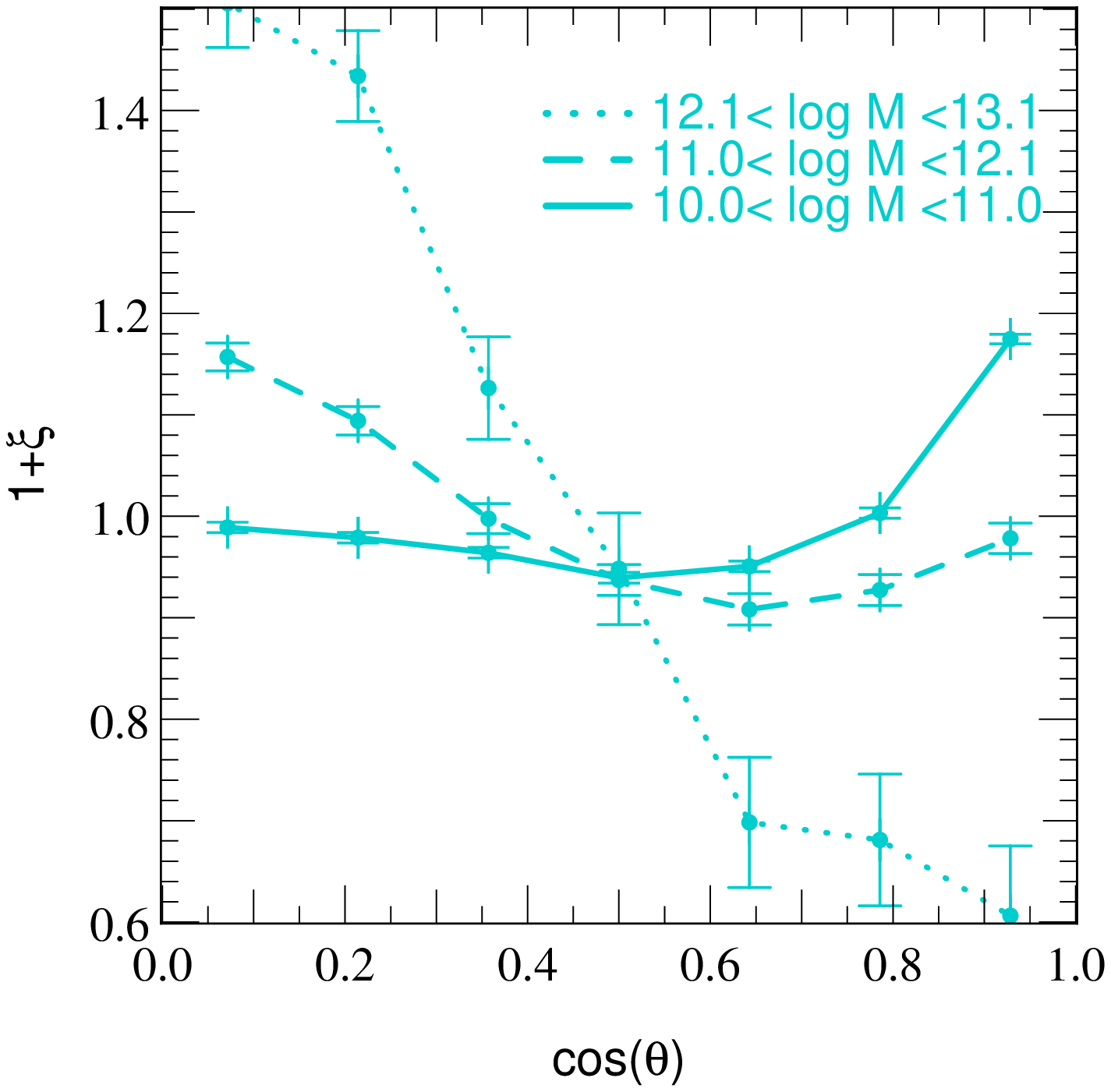}
\includegraphics[width=0.66\columnwidth]{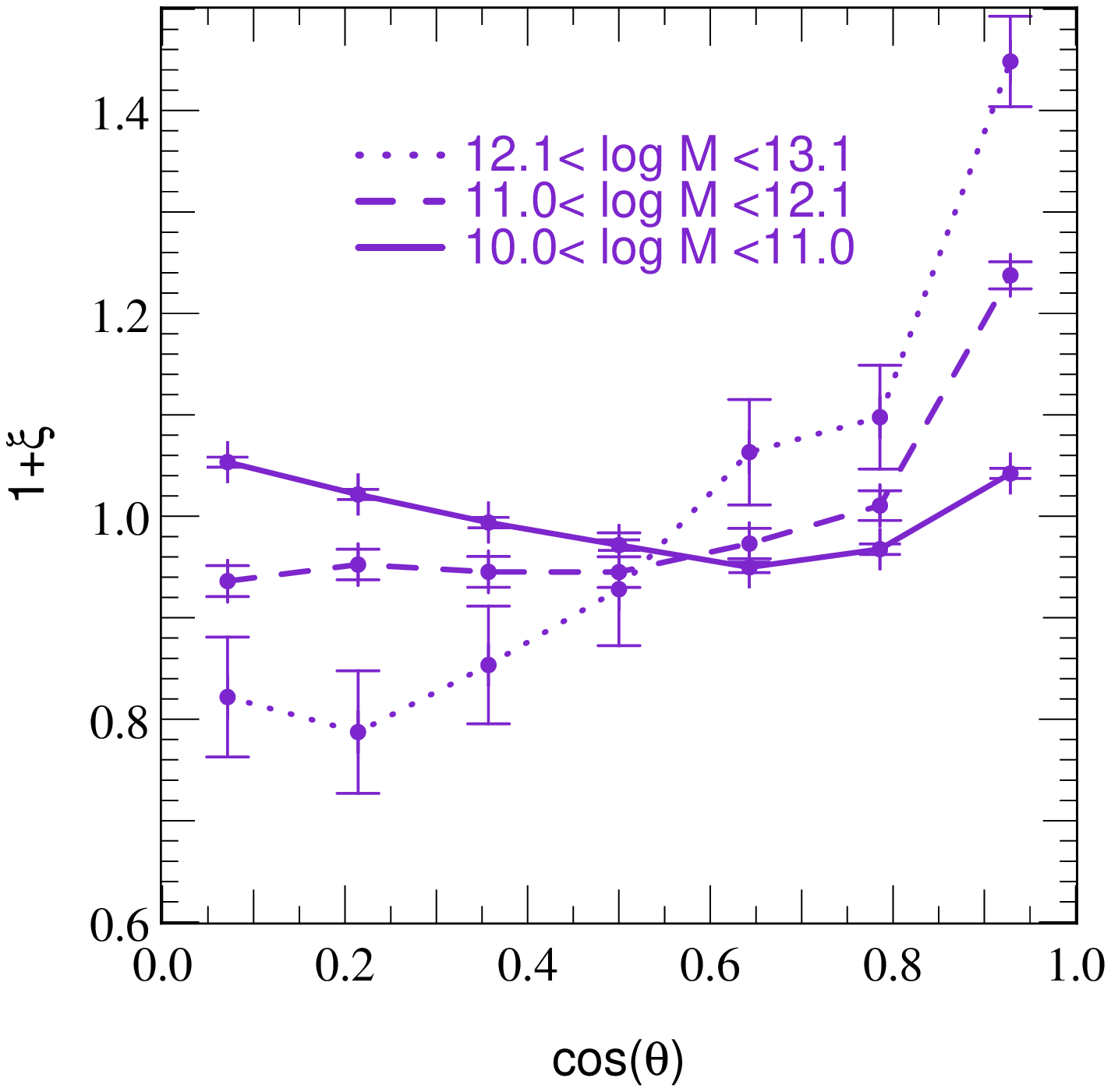}
\includegraphics[width=0.66\columnwidth]{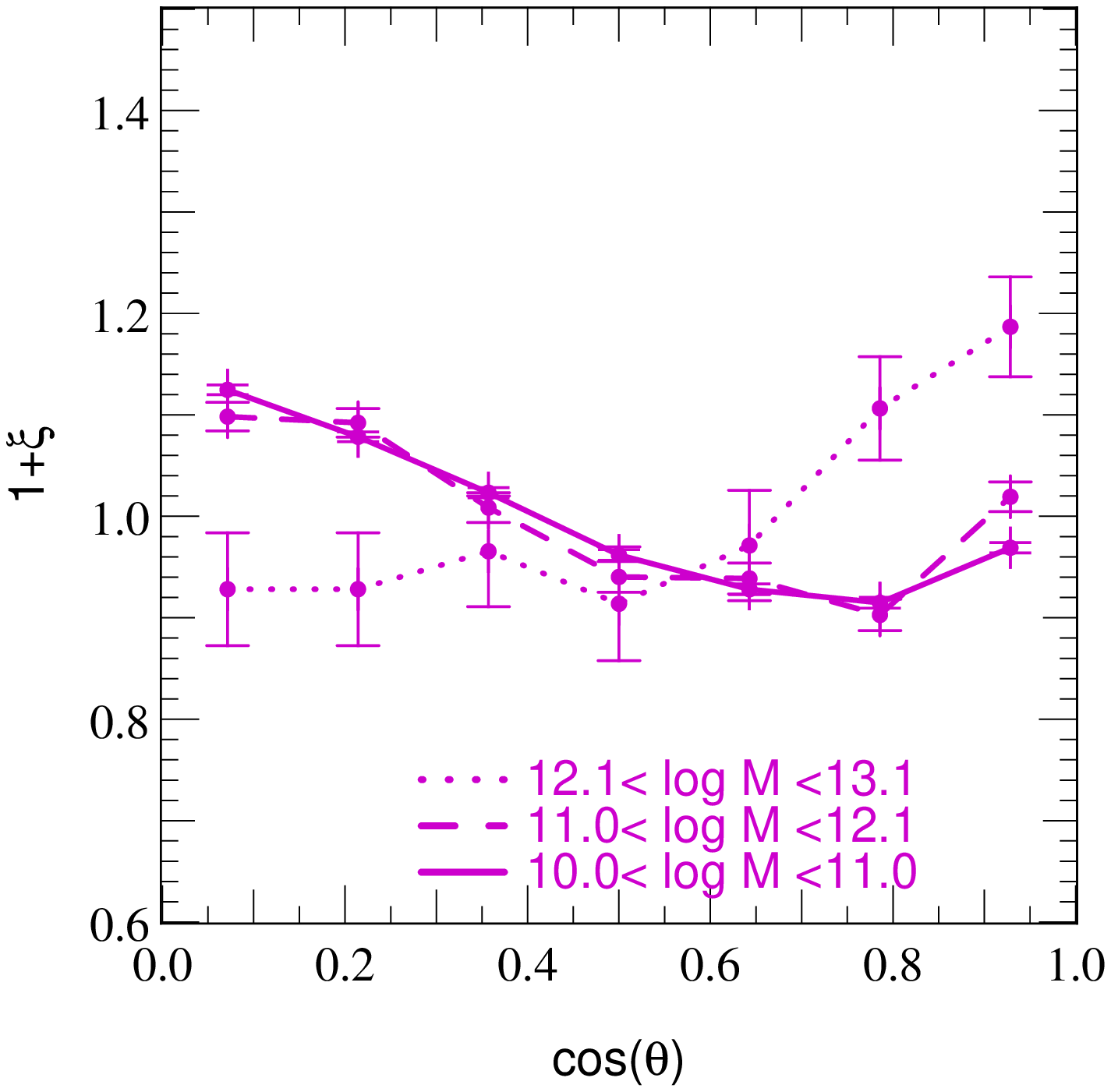}
\caption{Same as Fig.~\ref{fig:tidaltensor-gal} for DM halos in the \hagn simulation. The three different axis of the tidal tensor are colour-coded from cyan ($\mathbf{e}_{1}$) to magenta ($\mathbf{e}_{3}$) through purple ($\mathbf{e}_{2}$).
Different mass bins are coded from solid ($10^{10}$ to $10^{11}\, \rm M_\odot$) to dotted  ($10^{12}$ to $10^{13}\, \rm M_\odot$) through dashed lines ($10^{11}$ to $10^{12}\, \rm M_\odot$).
A transition is detected: the spin of high-mass halos tend to be aligned with the
intermediate 
({center purple panel}) and with less probability major (right magenta panel) principal axis, whereas the spin of  low-mass halos is
more likely to point along the minor axis (left cyan panel). 
Those findings are consistent with previous studies of dark matter only simulations.
\label{fig:tidaltensor-DM} 
}
\end{figure*}

Fig.~\ref{fig:tidaltensor-DM} displays the PDF of the cosine of the angle between the tidal eigen-directions and the spin of dark matter halos of different masses. The same qualitative behaviour as for galaxies is detected: the less massive halos have a spin preferentially aligned with $\mathbf{e}_{1}$ (somehow the filaments) while higher masses have their spin perpendicular to it (in agreement with what previous studies found in dark matter only simulations).

\begin{figure}
\includegraphics[width=\columnwidth]{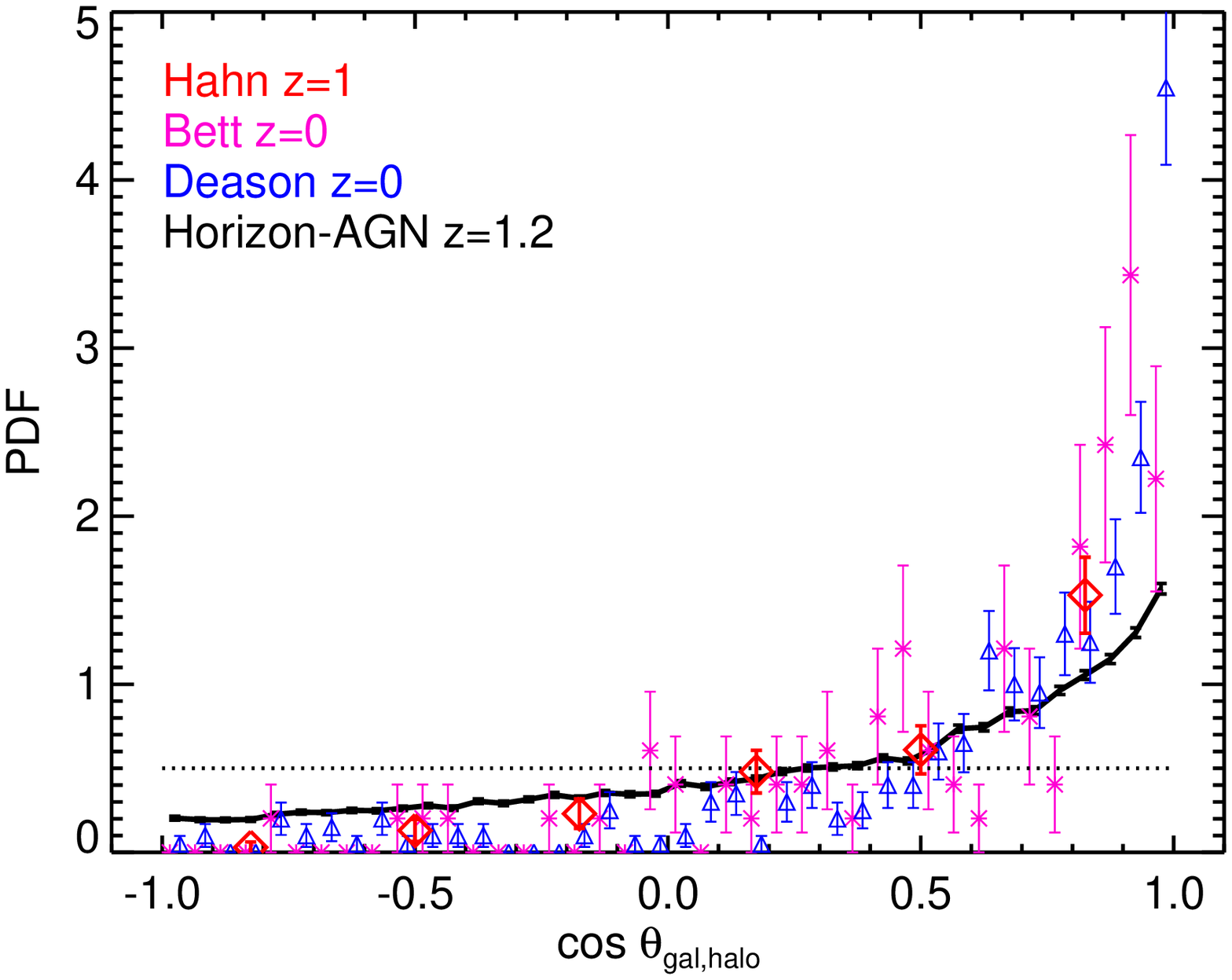}
\caption{PDF of the cosine of the angle between the spin of the DM halo and that of the galaxy in our simulation at $z=1.2$ (solid) and from~\citet{hahn10} at $z=1$ (red diamonds), from~\citet{Bett10} (pink stars) and from~\citet{Deason11} (blue triangles) both at $z=0$. The dotted line refers to a uniform isotropic distribution. Error bars are the 1-$\sigma$ standard deviation for Poisson statistics.}
\label{fig:spin-gal-halo} 
\end{figure}

Fig.~\ref{fig:spin-gal-halo} shows the PDF of the cosine of the angle between the spin of the DM halo and the spin of the galaxy for $\sim 32\, 000$ pairs at $z=1.2$ for halos with virial mass $M_{\rm h}> 10^{11}\,Ê\rm M_\odot$. We compare our result to that of~\cite{hahn10} ($z=1$, $M_{\rm h}> 10^{11}\,Ê\rm M_\odot$, 89 pairs),~\cite{Bett10} ($z=0$, $M_{\rm h}>5\times 10^{9}\,Ê\rm M_\odot$, 99 pairs) and~\cite{Deason11}  ($z=0$, $5\times 10^{11}<M_{\rm h}<5\times 10^{12}\,Ê\rm M_\odot$, 431 pairs). It shows that galaxies are slightly less aligned with their host DM halo in our \hagn simulation than what is found by other works, and used in the semi-analytic model of~\cite{Joa++13b} (based on~\citealp{Bett10} and \citealp{Deason11}), even though all results are compatible within 2-$\sigma$ error bars.
This slight difference might originate from AMR (\hagn and~\citeauthor{hahn10}) versus smoothed particle hydrodynamics (SPH,~\citeauthor{Bett10} and~\citeauthor{Deason11}), the persistence of cold flows~\citep{nelsonetal13}, disc versus elliptical galaxies~\citep{scannapiecoetal2012}, representativity of their re-simulations, etc.
Note that both AMR runs (\hagn and that of~\citeauthor{hahn10} performed with the {\sc ramses} code as well) at the same redshift are in better agreement.

\end{document}